\documentclass{aa}
\usepackage[utf8]{inputenc}
\usepackage{graphicx}
\usepackage[english]{babel}
\usepackage{amsmath}
\usepackage{mathtools}
\usepackage{amssymb}
\usepackage{txfonts}
\usepackage{hyperref} 
\usepackage{color}
\usepackage{rotating}
\usepackage{enumitem}
\usepackage[labelfont=bf]{caption}
\usepackage{subcaption}
\usepackage{ragged2e}
\usepackage{threeparttable}
\usepackage{placeins}
\usepackage{url}
\hypersetup{colorlinks=true,citecolor=blue}

\definecolor{orange}{rgb}{0.99,0.69,0.07}

\def\factor{\epsilon}

\newcommand{\eq}[1]{Eq.\,(\ref{#1})}
\newcommand{\eqs}[2]{Eqs.\,(\ref{#1}) and (\ref{#2})}
\newcommand{\eqsss}[4]{Eqs.\,(\ref{#1}), (\ref{#2}), (\ref{#3}), and (\ref{#4})}
\newcommand{\fig}[1]{Fig.\,\ref{#1}}
\newcommand{\figs}[2]{Figs.\,\ref{#1} and \ref{#2}}
\newcommand{\sect}[1]{Sect.\,\ref{#1}}
\newcommand{\sects}[2]{Sects \,\ref{#1} and \ref{#2}}

\newcommand{\app}[1]{Appendix\,\ref{#1}}
\newcommand{\tab}[1]{Table\,\ref{#1}}

\begin{document}

\title{How does the background atmosphere affect the onset of the runaway greenhouse?}
\author{Guillaume Chaverot\inst{1}, Martin Turbet\inst{1}, Emeline Bolmont\inst{1} and J\'er\'emy Leconte\inst{2}}

\offprints{G. Chaverot,\\ email: guillaume.chaverot@unige.ch}

\institute{
$^1$ Observatoire de Gen\`eve, Universit\'e de Gen\`eve, Chemin Pegasi 51b, CH-1290 Sauverny, Switzerland\\
$^2$ Laboratoire d'astrophysique de Bordeaux, Univ.  Bordeaux, CNRS, B18N, all\'ee Groffroy Saint-Hilaire, Pessac, F-33615, France}

  \date{Accepted to A\&A}
  \abstract
   {As the insolation of an Earth-like (exo)planet with a large amount of water increases,
   its surface and atmospheric temperatures also increase, eventually leading to a catastrophic runaway greenhouse transition. 
   While some studies have shown that the onset of the runaway greenhouse may be delayed due to an overshoot of the outgoing longwave radiation (OLR) - compared to the Simpson-Nakajima threshold - by radiatively inactive gases, there is still no consensus on whether this is occurring and why.
   Here, we used a suite of 1D radiative-convective models to study the runaway greenhouse transition, with particular emphasis on taking into account the radical change in the amount of water vapour (from trace gas to dominant gas).  
   The aim of this work is twofold: first, to determine the most important physical processes and parametrisations affecting the OLR; and second, to propose reference OLR curves for N$_2$+H$_2$O atmospheres. 
   Through multiple sensitivity tests, we list and select the main important physical processes and parametrisations that need to be accounted for in 1D radiative-convective models to compute an accurate estimate of the OLR for N$_2$+H$_2$O atmospheres. The reference OLR curve is computed with a 1D model built according to the sensitivity tests. These tests also allow us to interpret the diversity of results already published in the literature. Moreover, we provide a correlated-k table able to reproduce line-by-line calculations with high accuracy.
   We find that the transition between an N$_2$ -dominated atmosphere and an H$_2$O-dominated atmosphere induces an overshoot of the OLR compared to the (pure H$_2$O) Simpson–Nakajima asymptotic limit. This overshoot is first due to a transition between foreign and self-broadening of the water absorption lines, and second to a transition between dry and moist adiabatic lapse rates.}
  

\titlerunning{Background atmosphere effect on the onset of the runaway greenhouse}
\authorrunning{Chaverot, Turbet, Bolmont \& Leconte} 

   \maketitle

\section{Introduction}

Terrestrial planets can retain liquid water at their surface when the absorbed stellar radiation is balanced by the thermal emitted flux. If this balance is broken and if the absorbed flux is the highest, the planet warms to reach a new stable state at a higher temperature. 
Due to the strong opacity of water in the infrared, the thermal emission can be fully absorbed by the atmosphere when the temperature is high enough. 
Consequently, the thermal emission reaches a maximum, named the Simpson-Nakajima limit \citep{simpson_studies_1929, nakajima_study_1992}. Therefore, the planet stays in an unstable state and a catastrophic positive feedback arises and dramatically increases its temperature. 

\begin{figure*}[h!]
    \centering\includegraphics[width=\linewidth]{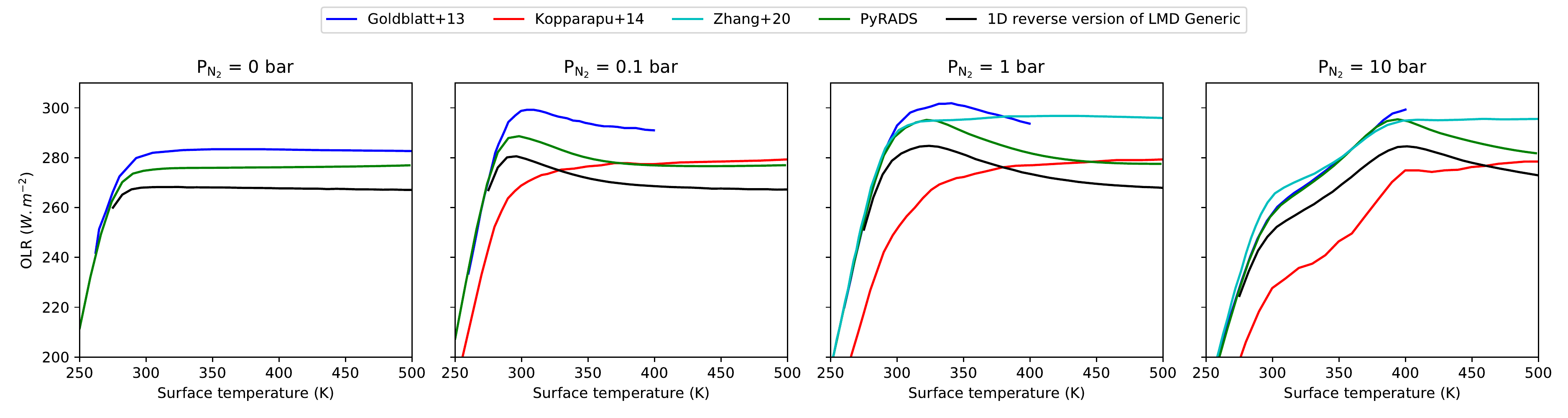}
    \caption{OLR as a function of surface temperature for different N$_2$ partial pressures (P$_{\textrm{N}_2}=\{0, 0.1,1,10\}$~bar) found in the literature \citep{kopparapu_habitable_2014,goldblatt_low_2013, zhang_how_2020} or computed in this work using the following existing models: the 1D reverse version of LMD-Generic (also known as kcm1d; \citep{turbet_runaway_2019} and PyRADS \citep{koll_earths_2018}.
    \cite{kopparapu_habitable_2014} also added 350~ppm of CO$_2$ to the radiative transfer calculation (Ravi K. Kopparapu, personal communication).}
    \label{model_comparison}
\end{figure*}

There is a strong interest to study this runaway greenhouse effect \citep{komabayasi_discrete_1967, ingersoll_runaway_1969, kasting_runaway_1988, nakajima_study_1992, goldblatt_runaway_2012,boukrouche_beyond_2021}, in particular to accurately determine the Simpson-Nakajima limit. This would allow us to know how close the Earth is to the runaway greenhouse threshold, but also to determine the inner edge of the habitable zone (HZ) more precisely \citep[e.g.][]{kasting_habitable_1993, kopparapu_habitable_2013, zhang_how_2020}. 
Some studies \citep[e.g.][]{goldblatt_low_2013, kopparapu_habitable_2014, ramirez_effect_2020} have shown that the outgoing longwave radiation (OLR) may be strongly modified by radiatively inactive gases (e.g. N$_2$ or O$_2$, as in the Earth's atmosphere). Other studies \citep{pierrehumbert_principles_2010, koll_hot_2019} have shown that such gases may lead to an overshoot of the Simpson-Nakajima limit, thus delaying the runaway greenhouse positive feedback.
These OLR values, overshot compared to the pure water runaway greenhouse, are interpreted in \cite{pierrehumbert_principles_2010} as a coupled effect of the pressure broadening -- which increases the absorption -- and a lapse rate -- which shifts towards a dry adiabatic lapse rate, thus reducing the absorption.
\cite{kopparapu_habitable_2013} also suggested that the differences could come from the parametrisation of the adiabatic lapse rate. \cite{kopparapu_habitable_2013} used the \cite{kasting_runaway_1988} formulation, while other papers used the \cite{ding_convection_2016} formulation. Moreover, \cite{koll_hot_2019} interpreted the OLR overshoot as a modification of the scale height of the atmosphere due to a change of the mean molecular weight.
Possible interpretations are multiple, but there is still no consensus in the literature as of whether an OLR overshoot is really expected or not (see \fig{model_comparison}). The Simpson–Nakajima limit is quite well constrained for a pure vapour atmosphere (see \fig{model_comparison}, first panel), but differences between the various studies appear even with 0.1~bar of N$_2$, and reach up to 35~W.m$^{-2}$ for 10~bar of N$_2$. 
These discrepancies increase with the nitrogen pressure. Thus, at least one important process may be unknown - or not well constrained - to accurately compute the OLR of such atmospheres. 

The first aim of this work is to determine, through sensitivity tests, the main physical processes and parametrisations involved in the computation of the OLR with a 1D radiative-convective model.
This can be useful to better constrain 3D GCM simulations through a better understanding of some of the main physical processes that may be at play.
Secondly, we propose a reference OLR curve (\fig{fig_PyRADS-Conv1D}, top panel), done with a model built according to the sensitivity tests, for a H$_2$O+N$_2$ atmosphere, and to solve the question of the potential overshoot. We also propose a H$_2$O+N$_2$ correlated-k table, which provides similar results as a line-by-line computation.

In this work, we used a suite of 1D radiative-convective models already described in the literature. These models, presented in \tab{table_models} (see also \sect{sub_models}), are composed of a convection scheme and a radiative transfer model.
For this reason, in the method section (\sect{sec_method}) we firstly introduce different convection schemes (\sect{sub_conv_schemes}) and radiative transfer methods (\sect{sub_RT}), and finally we present the set of models we used (\sect{sub_models}). We present our results in \sect{sec_results} and we discuss them in \sect{sec_discussion}. 
We provide our conclusions in \sect{sec_conclusion}.

\section{Method}
\label{sec_method}
With the aim of reconciling the large range of results proposed in the literature (\fig{model_comparison}), we explored physical mechanisms induced by the addition of nitrogen in a pure steam atmosphere. We built a line-by-line 1D radiative-convective model usable as reference model, named \textit{PyRADS-Conv1D}, to compute the OLR of the atmosphere. 
The radiative transfer calculation is derived from the PyRADS\footnote{\url{https://github.com/ddbkoll/PyRADS}} model \citep{koll_earths_2018}, and the convective scheme is derived from the RADCONV1D model \citep{marcq_thermal_2017}.

To be confident with the physical processes to be included in PyRADS-Conv1D, we performed multiple sensitivity tests using different convective schemes (\sect{sub_conv_schemes}) and radiative transfer methods (\sect{sub_RT}) from 1D models already described in the literature and presented in Table~\ref{table_models}.
An exhaustive list of these tested processes and parametrisations is presented below (see also Table~\ref{table_tests}). The complete overview of the impact of each parametrisation on the OLR value is available in \app{sub_sensitivity}.
A similar comparison was done by \cite{yang_differences_2016} with other 1D models, and they found the same wide range of results. More precisely, they show that differences appear using similar models and they explain that these differences are mainly induced by the choice of the water continuum. Our more detailed sensitivity tests were motivated by the interesting conclusions of this study. We tested the following parametrisations: 1) the convection scheme, 2) the resolution of the vertical grid, 3) the shape of the absorption lines, 4) the spectral resolution of the absorption spectrum, 5) a self- of foreign broadening of the absorption lines, 6) the quantity of water isotopes, 7) the truncation of the water continuum, 8) with or without the H$_2$O-N$_2$ continuum, 9) the version of the H$_2$O-H$_2$O and H$_2$O-N$_2$ continua, 10) with or without the N$_2$-N$_2$ Collision Induced Absorption (CIA) continuum, 11) the effect a additional fixed amount of CO$_2$, 12) the solution of the two-stream equations, 13) the interpolation scheme of the correlated-k method. 


\subsection{Convection schemes}
\label{sub_conv_schemes}

The atmospheres we studied are made of a variable amount of condensable gas (water) and a fixed amount of background gas (nitrogen). 
The quantity of each gas in the atmospheric layer is dictated by the surface temperature and the surface partial pressures and following a moist adiabatic lapse rate (hereafter `moist adiabat').
Generally, we assume that condensates are immediately removed by precipitation. This pseudo-adiabatic hypothesis is motivated by the fact that 1D models lack a self-consistently calculated description of (inherently 3D) cloud formation and thus precipitation. Consequently, these models are considered cloud free without other assumptions. 
The atmospheric profiles are assumed to be fully saturated, and thus the pressure of the condensable gas (here water) is equal to the saturation pressure. In other words, the relative humidity (RH) is equal to unity except in the stratosphere.

The main moist adiabats provided by (and used in) the literature are proposed by \cite{kasting_runaway_1988} and \cite{ding_convection_2016}. An analytic comparison is made in \app{sub_adiabat}, and it highlights that the main difference between them is the assumption used to define the entropy of the condensable gas. 
With the aim of comparing and discussing the assumptions and parametrisations done to compute the atmospheric profiles of 1D radiative-convective models, we used two different convective schemes in our
study, which are based on the two different adiabats of the literature. 
The first one, Conv\_K88, is based on the adiabat proposed in \cite{kasting_runaway_1988}, and the second one, Conv\_D16, uses the adiabat proposed in \cite{ding_convection_2016}.
These schemes also assume two different definitions of the vertical grid (see \app{sub_vertical_grid}). A complete description of these schemes is given here.

The temperature profiles made by the \textit{Conv\_K88} scheme follow a lapse rate computed using the equation from \cite{kasting_runaway_1988} with 200 atmospheric levels and a minimum total pressure at the top fixed at 0.1~Pa. The saturation pressure and the water entropy are computed using experimental look-up tables \citep{haar_nbsnrc_1984}. 
When the temperature reaches 200~K, the atmosphere is considered isothermal with a constant water-mixing ratio \citep{kasting_habitable_1993}. The change of gravity in the atmosphere due to the altitude is taken into account. 
This method was developed for RADCONV1D \citep{marcq_thermal_2017}. 

The \textit{Conv\_D16} scheme uses the \cite{ding_convection_2016} adiabatic lapse rate, with 100 levels and a fixed pressure difference between the top and the bottom of the atmosphere (10$^6$~Pa). The pressure of the highest level is not fixed, but the two previous parameters are interdependent and chosen to correctly represent the whole atmosphere (see \app{sub_vertical_grid} for more details). The saturation pressure is computed using the Clausius-Clapeyron equation. By definition of the \cite{ding_convection_2016} formulation of the adiabatic lapse rate, the water entropy is computed using the perfect gas approximation (see \app{sub_adiabat}). When the temperature reaches 150~K, the atmosphere is considered isothermal and the water-mixing ratio is fixed. 
This method was developed for PyRADS and the original parameters are available in \cite{koll_earths_2018}.

\begin{figure}[b!]
    \centering\includegraphics[width=\linewidth]{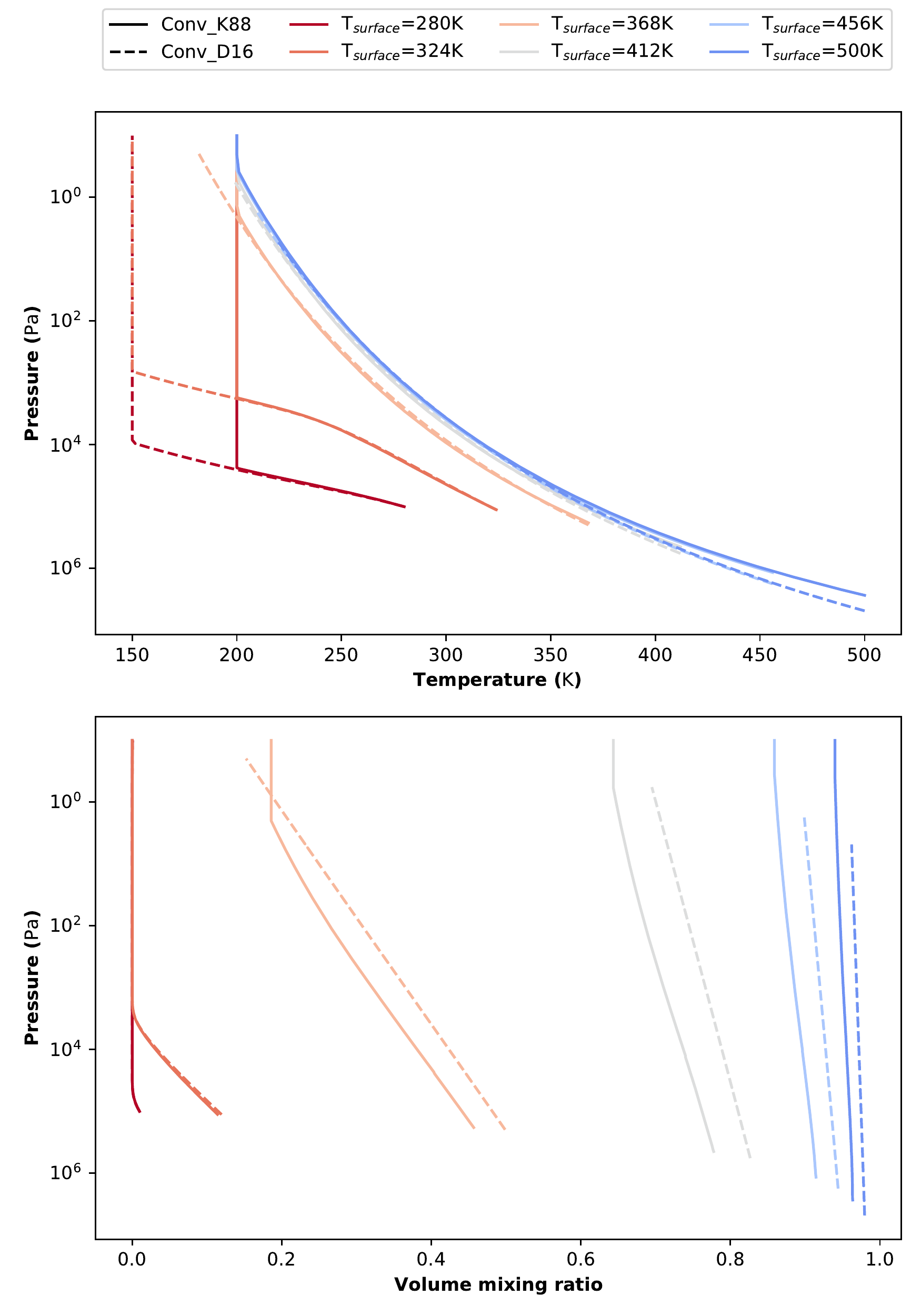}
    \caption{Comparison between Conv\_K88 (full lines) and Conv\_D16 (dotted lines) convection schemes for different surface temperatures. The top panel represents the temperature profiles, and the bottom panel represents the volume-mixing ratio of the water for different surface temperatures. Here, P$_{\textrm{N}_2}$=1~bar.}
    \label{fig_adiabat}
\end{figure}

We fixed the number of atmospheric levels of each method thanks to a test of convergence of the OLR value relative to the vertical resolution (see \app{sub_vertical_grid}).
Figure~\ref{fig_adiabat} highlights differences of the thermal profiles and of the water-mass-mixing ratio profiles computed using one convection scheme or the other. The Conv\_K88 scheme is more accurate because it is based on look-up tables, while the Conv\_D16 scheme is based on perfect gas equations (e.g. Clausius-Clapeyron equation). 
This induces a non-negligible difference on the estimation of the OLR value (\fig{fig_conv}). For this reason, we used the Conv\_K88 scheme as a reference convective scheme (see \sect{sub_models}).
Figure~\ref{fig_adiabat} also shows that for a high surface temperature, that is a high surface pressure, the Conv\_D16 scheme is not able to represent the stratosphere because of the fixed difference of pressure between the top and the bottom of the atmosphere.
For both methods, we considered an Earth-like planet with a gravity acceleration at the surface equal to 9.81~m.s$^{-1}$.

It is important to keep in mind that for both convection schemes presented in this work the pseudo-adiabatic hypothesis induces a water pressure equal to the saturation pressure (i.e. given by the temperature).
Therefore, as the surface boundary conditions are fixed (temperature and nitrogen pressure), the surface water pressure stays unchanged whatever the background gas pressure is.

\begin{table*}[ht]\footnotesize\centering
\setlength{\doublerulesep}{\arrayrulewidth}
\captionsetup{justification=justified}
\caption{Main characteristics of 1D radiative-convective models used in this work. The models kcm1d, PyRADS, and \texttt{Exo\_k} are available in the literature while PyRADS-Conv1D is an original model. In bold, we give the models used to produce reference curves using the line-by-line or the correlated-k radiative transfer methods.}
\label{table_models}
\begin{tabular}[c]{c c c c}
    \multicolumn{4}{c}{\textbf{Models overview}} \\
    \hline\hline\hline
    Models & Convection method & Radiative transfer method & Ref.\\
    \hline
    kcm1d & Conv\_K88 & kcm1d-RT & \cite{turbet_runaway_2019} \\
    \textbf{\texttt{Exo\_k}} & Conv\_K88 & \texttt{Exo\_k}-RT & \cite{leconte_spectral_2021} \\
    PyRADS & Conv\_D16 & PyRADS-RT & \cite{koll_earths_2018} \\
    \textbf{PyRADS-Conv1D} & Conv\_K88 & PyRADS-updated & this work \\
    \hline\hline\hline
\end{tabular}\vspace{0.3cm}

\begin{tabular}[c]{c c c c c}
    \multicolumn{5}{c}{\textbf{Convection schemes}} \\
    \hline\hline\hline
    Conv. methods & Adiabatic lapse rate & Levels & Pres. boundary conditions & Entropy def. \\
    \hline
    Conv\_D16 & \cite{ding_convection_2016} & 100 & fixed diff. top and bottom & perfect gas \\
    Conv\_K88 & \cite{kasting_runaway_1988} & 200 & fixed top pressure & experiments \\
    \hline\hline\hline
\end{tabular}\vspace{0.3cm}

\begin{tabular}[c]{ c c c c c c c }
    \multicolumn{7}{c}{\textbf{Radiative transfer methods}} \\
     \hline\hline\hline
    R.T. methods & Absorb. Coeff. & Line shape & HITRAN &  H$_2$O Cont. & N$_2$-N$_2$ Cont. & H$_2$O iso. \\
    \hline
    kcm1d-RT & correlated-k & Voigt (high res.) & 2016 & MT\_CKD2.5 - [0.1, 10k]cm$^{-1}$ & HITRAN CIA & yes \\
    \texttt{Exo\_k}-RT & correlated-k/cross section & Voigt (high res.) & 2016 & MT\_CKD3.2 - [0.1, 20k]cm$^{-1}$ & HITRAN CIA & yes \\
    PyRADS-RT & line-by-line (0.01 cm$^{-1}$) & Lorentz & 2016 & MT\_CKD3.2 - [0.1, 10k]cm$^{-1}$ & without & no \\
    PyRADS-updated & line-by-line (0.01 cm$^{-1}$) & Lorentz & 2016 & MT\_CKD3.2 - [0.1, 20k]cm$^{-1}$ & HITRAN CIA & no \\
    \hline\hline\hline
\end{tabular}\normalsize
\vspace{0.2cm}
\justifying
\textbf{Notes.} Convection scheme and the radiative transfer method of each model are detailed in the corresponding tables. The convection scheme sub-table includes the adiabatic lapse rate used (Adiabatic lapse rate), the number of levels in the atmosphere (Levels), the pressure boundary conditions at the top and at the bottom of the atmosphere (Pres. boundary conditions), and the assumption used to define the entropy of the condensable gas (Entropy def.). The radiative transfer method table includes the method to compute absorption coefficients (Absorb. Coeff.), the shape of the absorption lines (Line shape), the database used to compute the absorption lines (HITRAN) \citep{gordon_hitran2016_2017-1}, the database of the H$_2$O-H$_2$O and H$_2$O-N$_2$ continua (H$_2$O Cont.) \citep{amundsen_treatment_2017}, the database of the N$_2$-N$_2$ continuum (N$_2$-N$_2$ Cont.) \citep{karman_update_2019}, and if the method considers the water isotopes lines from HITRAN database at terrestrial abundances \citep{de_bievre_isotopic_1984} (H$_2$O iso.).\\

\end{table*}
\subsection{Radiative transfer methods}
\label{sub_RT}

In this work, we used different radiative transfer methods from four different models presented in Table \ref{table_models}. These methods are sub-divided between the two families of radiative transfer calculations detailed below: line-by-line and correlated-k. 

PyRADS-RT and PyRADS-updated are line-by-line methods, while kcm1d-RT and \texttt{Exo\_k}-RT are correlated-k methods.
Our aim here is to use a line-by-line method, through the PyRADS-Conv1D model, to produce a result as accurate as possible to benchmark the calculations performed using a correlated-k method, which is computationally much more efficient. We used \texttt{Exo\_k} as reference model for the correlated-k method.
This approach is motivated by the large dispersion of results in the literature and to reconcile line-by-line and correlated-k studies, which seemingly lead to different results (\fig{model_comparison}).

The specificities of the line-by-line and correlated-k methods are detailed below. Several sensitivity tests were performed to quantify the influence of each of the assumptions usually done in the literature on the final estimation of the OLR (\app{sub_sensitivity}). 

As shown in \tab{table_models}, for all methods the H$_2$O-H$_2$O and H$_2$O-N$_2$ continua are taken into account using the MT\_CKD database \citep{mlawer_development_2012}. The N$_2$-N$_2$ collision-induced absorption (CIA) is taken from the HITRAN CIA database \citep{karman_update_2019}.
According to the MT\_CKD formalism \citep{mlawer_development_2012} the water vapour absorption lines are truncated at 25~cm$^{-1}$ and the plinth\footnote{For each spectral line $\nu$, the MT\_CKD continuum formalism assumes a constant absorption value between $\mathrm{\nu - 25~cm^{-1}}$ and $\mathrm{\nu + 25~cm^{-1}}$ equal to the absorption at $\mathrm{\nu \pm 25~cm^{-1}}$. Consequently this `plinth' needs to be removed in the line centre calculation.} is removed from the line centre calculation. 
In PyRADS-updated, we use the MT\_CKD3.2 continuum as explained in \app{sub_continuum}.
We choose to include only the main isotope water absorption lines in the radiative transfer calculation of PyRADS-updated to increase the computation speed.
This is also motivated by our of knowledge of the typical isotopic fraction of the water on terrestrial exoplanets.
This assumption does not significantly change the final result (see \app{sub_isotopes}). 
Finally, to compute the pressure broadening of the lines we need to have the exponents that describe the temperature dependence for N$_2$ and H$_2$O (see \eq{eq_gamma}). The exponent for N$_2$ was taken equal to that of the (Earth) 'air' composition \citep{gordon_hitran2016_2017-1}. 
We do not include Rayleigh scattering by water vapour for either methods because it becomes important beyond 10000~cm$^{-1}$ \citep{kopparapu_habitable_2013} where the thermal emission (i.e. the Planck law) is null for the studied range of temperature.

The radiative methods presented here assume different approximations to solve the two-stream equations used to describe the radiative transport of an atmosphere through a upward and downward flux \citep{meador_two-stream_1980}. The kcm1d-RT method uses the classical hemispheric mean approximation while PyRADS-RT and PyRADS-updated and \texttt{Exo\_k}-RT assume $F^+=\pi I(\bar{\mu})$ with $\bar{\mu}=0.6$. This approximation is only valid in the context of a non-scattering atmopshere.

PyRADS-RT and PyRADS-updated assume $\bar{\mu}=0.6$ which yields less than 1\,W.m$^{-2}$ of difference compared to the modified hemispheric mean method (see \sect{sub_two-stream} for more details).

\subsubsection{Line-by-line method}
The most direct radiative transfer method is to compute the exact absorption spectrum at each level of the atmosphere, that is for each pressure to temperature couple. In other words, the line-by-line method computes the absorption at each wavelength while the correlated-k method considers a statistical distribution of the absorption spectrum.
This method avoids interpolations of spectra intrinsic to the correlated-k computation and allows greater flexibility relative to the composition of the atmosphere. 
However, it makes the computation more time-expensive. As line-by-line computations provide an exact result, we chose PyRADS-updated as a reference radiative transfer method.

The PyRADS-RT and PyRADS-updated methods (\tab{table_models}) use low-resolution spectra computed using the HITRAN 2016 line list \citep{gordon_hitran2016_2017-1} and assume the shape of the lines follows a Lorentz profile. We show in the Appendix that a spectral resolution of 0.01~cm$^{-1}$ is sufficient to keep a high accuracy, in agreement with \cite{koll_earths_2018}.
We adapted the PyRADS-RT method to create the PyRADS-updated method by adding the N$_2$-N$_2$ CIA computation from PyRADS-shortwave\footnote{\url{https://github.com/ddbkoll/PyRADS-shortwave}} \citep{koll_hot_2019} and using a version of MT\_CKD continuum extended up to 20000~cm$^{-1}$ (see \tab{table_models}). 

\subsubsection{Correlated-k method}
Models based on the correlated-k method \citep{fu_correlated_1992} are convenient to use because of their much shorter computation time. Here, absorption spectra are pre-computed for pressure, temperature, and mixing ratio reference values in a so-called correlated-k table. During the radiative transfer calculation, a statistical wavelength distribution of these spectra is interpolated \citep[e.g.][]{fu_correlated_1992, leconte_spectral_2021} on the pressure to temperature values of each atmospheric level.
The correlated-k method can in principle provide the same results as a line-by-line calculation, but at a much lower computational cost. To achieve this purpose the number of bands of the correlated-k table needs to be high enough, as well as the resolution of the pressure, temperature, and mixing ratio grids.
In this paper, we provide an H$_2$O+N$_2$ table that gives similar results to an exact calculation (\fig{fig_PyRADS-Conv1D}, top panel).\\
To build the correlated-k table used in this work, we computed a dataset of high-resolution spectra using the HITRAN 2016 database \citep{gordon_hitran2016_2017-1}. 
These spectra are calculated for multiple values of temperature (T=$\{50,110,170 \dots 710, 5000\}$~K), pressure (P=$\{0.1, 1, 10 \dots 10^{7}\}$~Pa), and water-mass-mixing ratio (Q=$\{10^{-6}, 10^{-5}, 10^{-4} \dots 1\}$). 
The 21 Gauss points of the table follow a Legendre distribution. The infrared (IR) band is defined in the [10; 4000]~cm$^{-1}$ range with a resolution R=8 (48 bands), which is sufficient to accurately compute water absorption spectra.
We also include absorption lines of the water isotopes at terrestrial abundances \citep{de_bievre_isotopic_1984} in the absorption spectra, but as shown in \app{sub_isotopes} this does not significantly change the OLR value. 
We share the correlated-k table used in this work\footnote{\url{https://zenodo.org/record/5359158}} for a larger spectral range ([0.1; 30000]~cm$^{-1}$) and a higher resolution (R=300). A usable table at lower resolution can easily be made by using the \texttt{Exo\_k} package\footnote{\url{http://perso.astrophy.u-bordeaux.fr/\~jleconte/exo\_k-doc/index.html}}.

\subsection{Description of the models}
\label{sub_models}

Every 1D radiative convective model presented in \tab{table_models} is composed of a convection scheme and a previously described radiative transfer method. The models used are presented here.

The 1D reverse version of LMD-Generic, also named kcm1d \citep{turbet_runaway_2019}, uses the Conv\_K88 scheme (developed for RADCONV1D; \citealt{marcq_thermal_2017}) and the kcm1d-RT radiative transfer method. This method, described above, is derived from the LMD-Generic model and includes some GCM-specific parametrisations (see  \sects{sub_two-stream}{sub_interpolation}).

The \texttt{Exo\_k} model \citep{leconte_spectral_2021} also uses the Conv\_K88 scheme. Its radiative transfer method, \texttt{Exo\_k}-RT is a correlated-k method that provides results in accordance with our reference model PyRADS-Conv1D (see top panel in \fig{fig_PyRADS-Conv1D}). For this reason, we use it as reference model for sensitivity tests related to the correlated-k method.

The PyRADS model \citep{koll_earths_2018} uses the Conv\_D16 scheme with the PyRADS-RT method. We adapted several parameters to fit our needs (e.g. the vertical grid, see \app{sub_vertical_grid}). The original parameters are available in \cite{koll_earths_2018}. 

We built the PyRADS-Conv1D model using the most accurate convection scheme (Conv\_K88) and radiative transfer method (PyRADS-updated) to produce reference OLR curves according to our sensitivity tests.

\section{Results}
\label{sec_results}
We present our results in this section. First, we describe the processes that induce a runaway greenhouse effect for a pure steam atmosphere (\sect{sub_runaway_water}). Secondly, we highlight the differences that arise by adding nitrogen and the processes that allow the OLR overshoot relative to the Simpson-Nakajima limit (\sect{sub_overshoot}). Finally, we discuss the two physical processes that induce the overshoot of the OLR in detail (\sects{sub_broad}{sub_mol_weight}).

\subsection{Runaway greenhouse of a pure water atmosphere}
\label{sub_runaway_water}

The global physical processes that induce a runaway greenhouse effect for a pure water atmosphere are now very well constrained \citep{ingersoll_runaway_1969, nakajima_study_1992, goldblatt_runaway_2012}, but numerical estimations of the value of the greenhouse asymptotic limit vary between the different studies, as shown in \fig{model_comparison}. 
These small differences may come from the assumptions made in models or from second-order neglected processes. To understand origins of these differences, we first describe the physical processes involved in a warming atmosphere. 

Consider an Earth-like rocky (exo)planet with a global ocean without background gases. At very low temperatures, typically lower than 290K, almost all water is condensed or solidified at the surface of the planet; thus, the (H$_2$O-dominated) atmosphere is very thin and the planet radiates similarly to a black body.
If the surface temperature increases, the thermal emission increases because of the Planck radiation law, and the surface water is expected to progressively evaporate. 
The thin water vapour atmosphere induces a greenhouse effect via IR absorption by water vapour, consequently the OLR is not strictly equal to the black body emission law. 
If the temperature increases again, the surface pressure reaches a limit, P$_{thick,}$ for which the atmosphere becomes optically thick at long wavelengths (i.e. in the IR absorption bands of water). The OLR is then decoupled from the surface temperature \citep{nakajima_study_1992,pierrehumbert_principles_2010,goldblatt_low_2013,boukrouche_beyond_2021}. 
As the temperature profile follows a purely wet adiabat, the structure of the atmosphere in the thin layers (i.e. `radiatively active layers' from the optically thick limit P$_{thick}$) up to the top stay unchanged by increasing the surface temperature (see blue curves, i.e. high temperatures, in \fig{fig_adiabat}).
Therefore, the absorption in these layers also remains unchanged.
For this reason, the OLR reaches an asymptotic limit known as the moist troposphere limit or the Simpson–Nakajima limit \citep{goldblatt_runaway_2012}. As a result, if the absorbed stellar radiation (ASR) is higher than this asymptotic limit, the radiative balance is broken and the surface temperature increases, inducing even more evaporation and progressively building up a thick water-steam-dominated atmosphere. 
This is the runaway greenhouse effect. When the entire ocean is evaporated and when the planet reaches extremely high temperatures of thousands of kelvins \citep[e.g.][]{kasting_habitable_1993,kopparapu_habitable_2013,turbet_runaway_2019}, the OLR increases again. For these temperatures, the planet radiates in the visible, where the water is optically thin.

\subsection{Runaway greenhouse of an H$_2$O+N$_2$ atmosphere: The overshoot effect}
\label{sub_overshoot}

\begin{figure}[h!]
\centering\includegraphics[width=\linewidth]{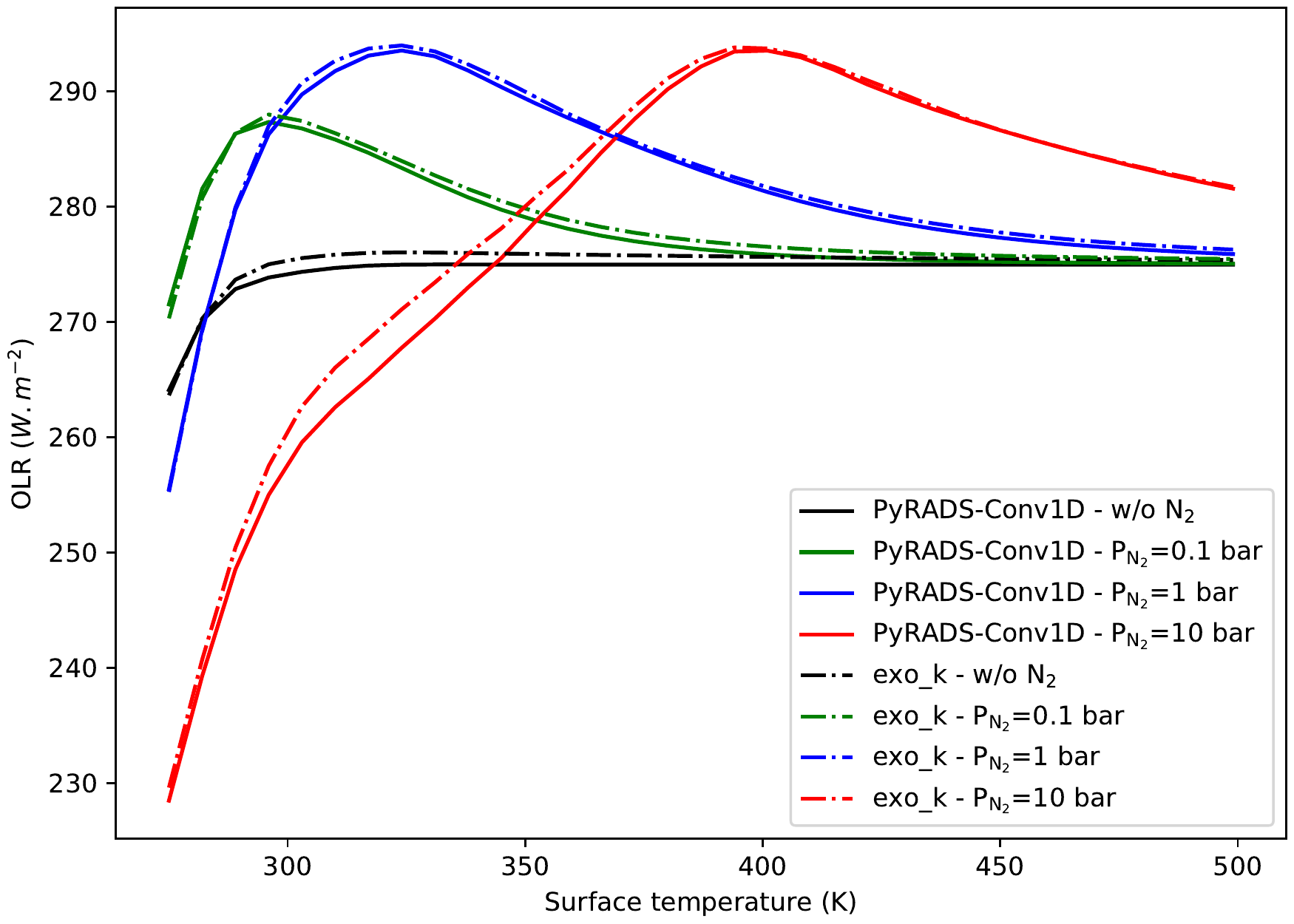}
\centering\includegraphics[width=\linewidth]{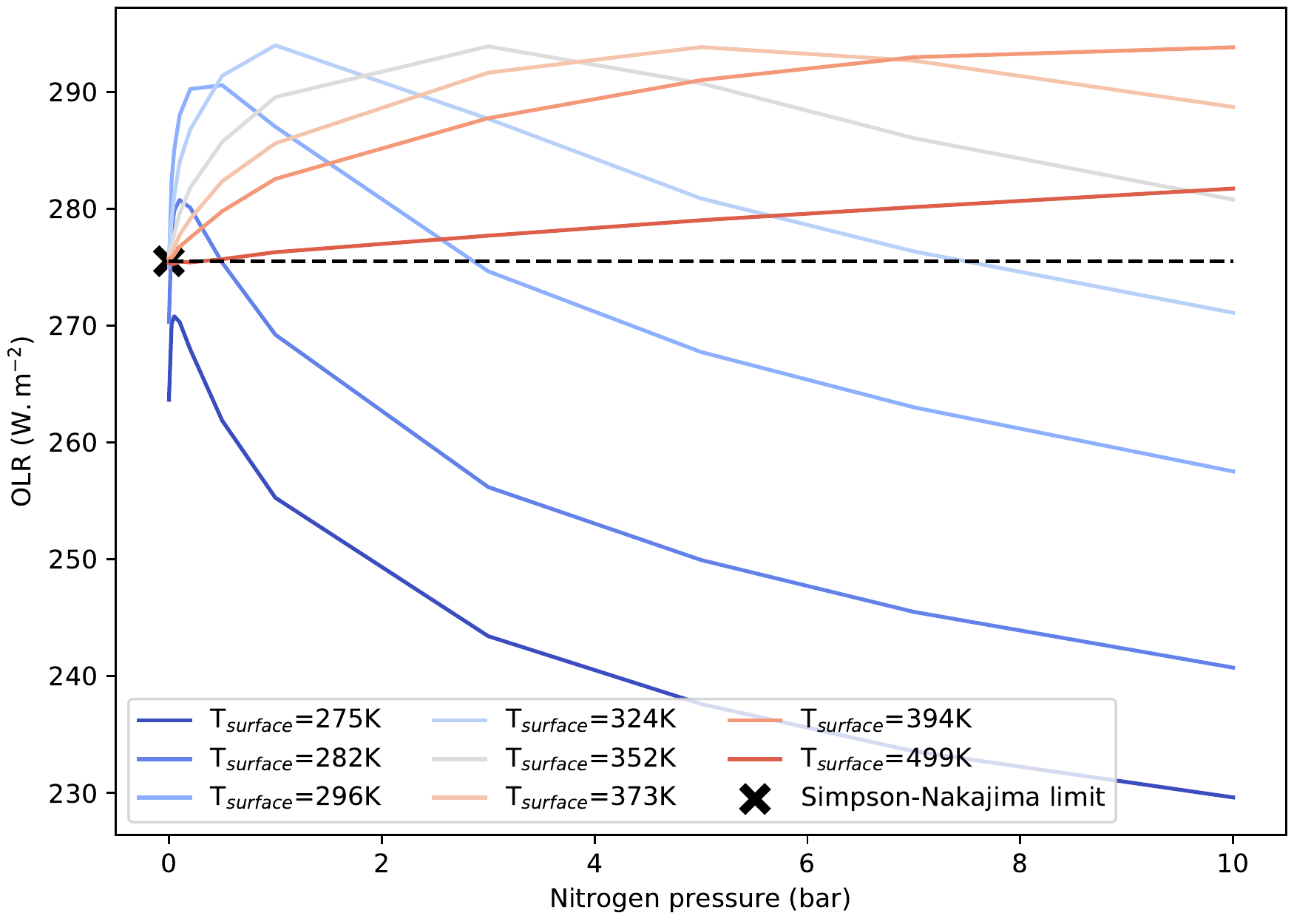}
\caption{OLR as a function of the surface temperature (top panel) or as a function of the nitrogen pressure (bottom panel).
The top panel represents the OLR curve for different N$_2$ pressures (black: P$_{\textrm{N}_2}$=0~bar, green: P$_{\textrm{N}_2}$=0.1~bar, blue: P$_{\textrm{N}_2}$=1~bar, red: P$_{\textrm{N}_2}$=10~bar) computed with PyRADS-Conv1D (full lines) and \texttt{Exo\_k} (dotted lines).
The bottom panel represents the OLR as a function of the surface nitrogen pressure for different fixed surface temperatures computed using \texttt{Exo\_k}. The black cross represents the Simpson-Nakajima limit for a pure vapour atmosphere.}
\label{fig_PyRADS-Conv1D}
\end{figure}

\begin{figure}[h!]
    \centering\includegraphics[width=\linewidth]{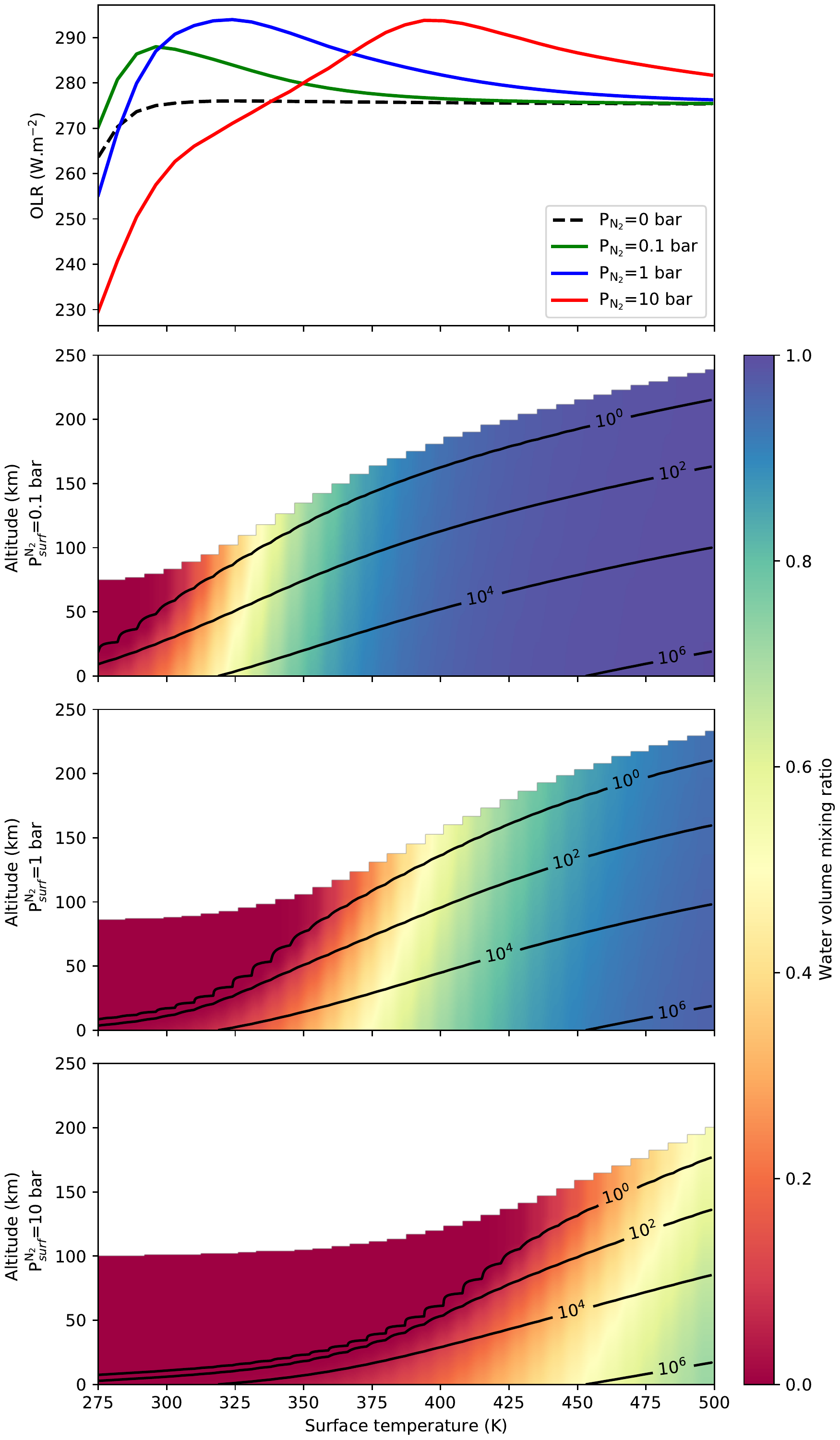}
    \caption{OLR and corresponding atmospheric water content as a function of the surface temperature. The top panel is the OLR as a function of the surface temperature for different nitrogen pressure, computed with PyRADS-Conv1D. The three bottom panels represent the size of the atmosphere (up to 0.1\,Pa) for different nitrogen pressures where the colour bar indicates the water-volume-mixing ratio ($\frac{P_{\mathrm{H_2O}}}{P_{\mathrm{H_2O}}+P_{\mathrm{N_2}}}$) of each atmospheric level. The contour lines indicate the partial pressure of the water in Pa units.}
    \label{fig_N2_mixing_ratio}
\end{figure}

In this section, we describe physical processes that may lead to or prevent an OLR overshoot, presented in the top panel of \fig{fig_PyRADS-Conv1D}, for an H$_2$O+N$_2$ atmosphere and computed with PyRADS-Conv1D and \texttt{Exo\_k}. 
Regarding \figs{fig_N2_mixing_ratio}{fig_OLRvsvmr}, the evolution of the OLR value is mainly determined by the water-volume-mixing ratio value, whatever the nitrogen pressure, according to three ranges of mixing ratio values: below $5\times10^{-5}$, between $5\times10^{-5}$ and 0.2, and above 0.2. Therefore, in the following we describe the evolution of the OLR through three cases: low, high, or intermediate mixing ratios.\\

\textbf{At very low mixing ratios,} the surface temperature is rather low (\fig{fig_N2_mixing_ratio}). Consequently, the vapour pressure is also low and N$_2$ is dominant; thus, the lapse rate is very close to a dry adiabatic lapse rate. For the following, we assumed a dry adiabat. Moreover, the broadening of the water absorption lines is dominated by foreign broadening (\fig{N2_broad}).

The atmospheric absorption is mainly in the first layers that contain water (with a partial water pressure roughly greater than 5~Pa). The temperature of a dry atmosphere is lower than for a wet one, that is the temperature decreases faster relatively to altitude (\fig{fig_adiabat}). 
Therefore, for a given low surface temperature, the total amount of steam in the whole atmosphere is low if the nitrogen pressure is high (because the lapse rate is close to a dry adiabatic lapse rate; see \fig{fig_N2_mixing_ratio}); thus, the intensity of absorption lines decreases by increasing the nitrogen pressure (e.g. bottom panel of \fig{fig_PyRADS-Conv1D} for T$_{surface}$=275\,K).
However, as the vapour pressure is limited by the pseudo-adiabatic hypothesis, it cannot exceed the saturation pressure (P$_{\textrm{H}_2\textrm{O}}\leq$ P$_{\textrm{saturation}}$); therefore, the self-broadening is limited as well. As nitrogen is a non-condensable gas here, its pressure is not limited and the foreign broadening may be stronger than the self-broadening. 
Therefore, when the atmopshere is nitrogen dominated, the total broadening of the absorption lines is stronger than for a pure vapour atmosphere, the atmosphere absorbs in more wavelengths, and the OLR is weaker than for a pure vapour atmosphere, as shown in the bottom panel of \fig{fig_PyRADS-Conv1D} and for the P$_{\textrm{N}_2}$=10~bar in the top panel of \fig{fig_PyRADS-Conv1D}.\\

\textbf{At high mixing ratios,} the temperature is high and water becomes dominant; hence, the temperature profile follows a moist adiabatic lapse rate. By evaporating more and more water, the absorption line broadening becomes dominated by the self-coefficient, and the mean molecular weight tends towards the water molecular weight. Properties of the atmosphere (temperature profile, mixing ratio, molecular weight, etc.) tend towards the properties of a pure water atmosphere; thus, the OLR tends towards the Simpson–Nakajima limit (\fig{fig_PyRADS-Conv1D}, bottom panel).\\

\begin{figure}[h!]
    \centering\includegraphics[width=\linewidth]{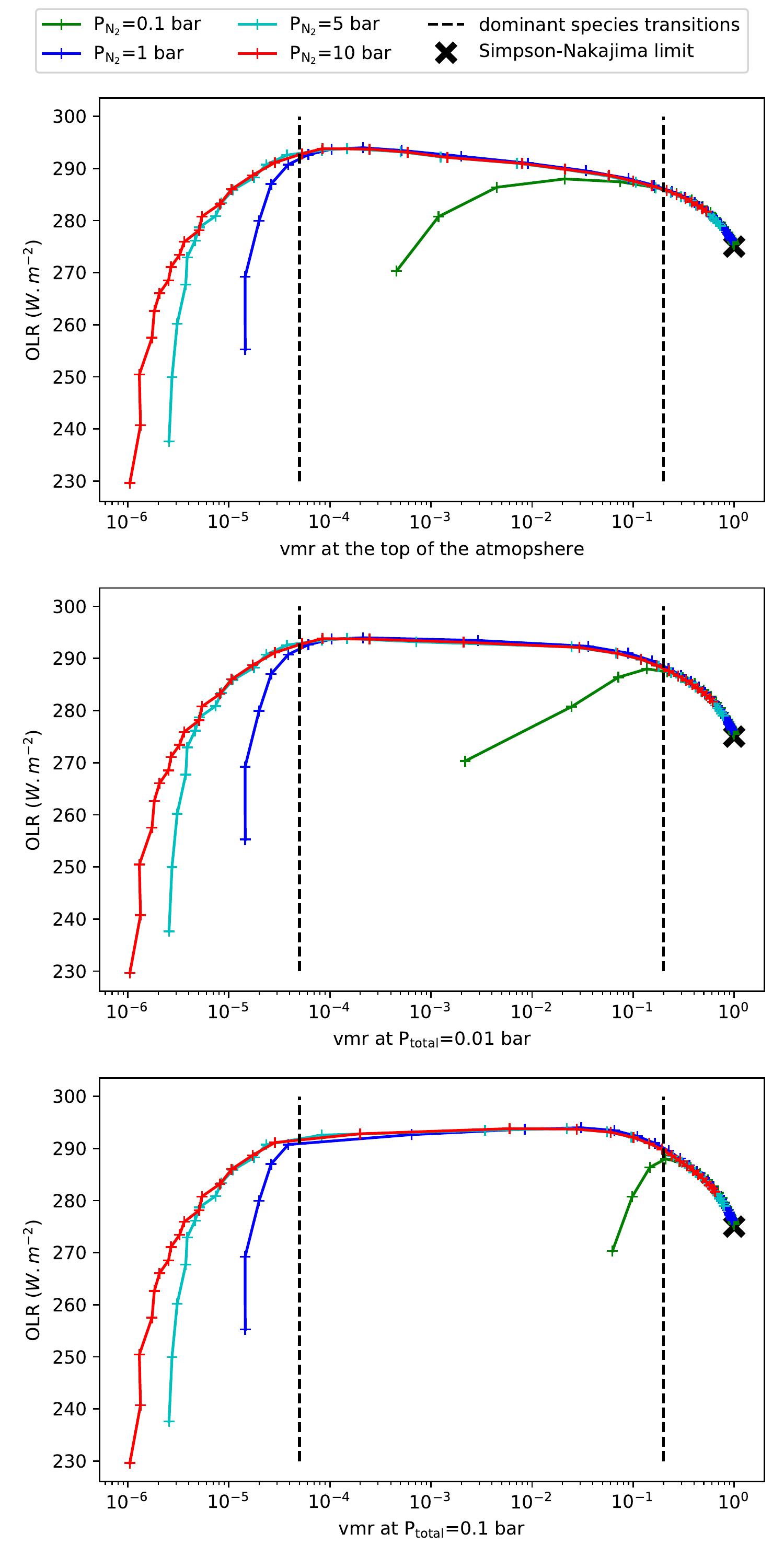}
    \caption{OLR as a function of the volume-mixing ratio for different nitrogen surface pressures using \texttt{Exo\_k}. The volume-mixing ratio (vmr) values are given at the top of the atmosphere (top panel), at 0.01\,bar (middle panel), and at 0.1\,bar (bottom panel) of total pressure. The black cross represent the Simpson-Nakajima limit for a pure vapour atmosphere, and the vertical dashed lines approximately delimit the range of vmr values for which there is a transition of dominant species.}
    \label{fig_OLRvsvmr}
\end{figure}

\begin{figure}[h!]
    \centering\includegraphics[width=\linewidth]{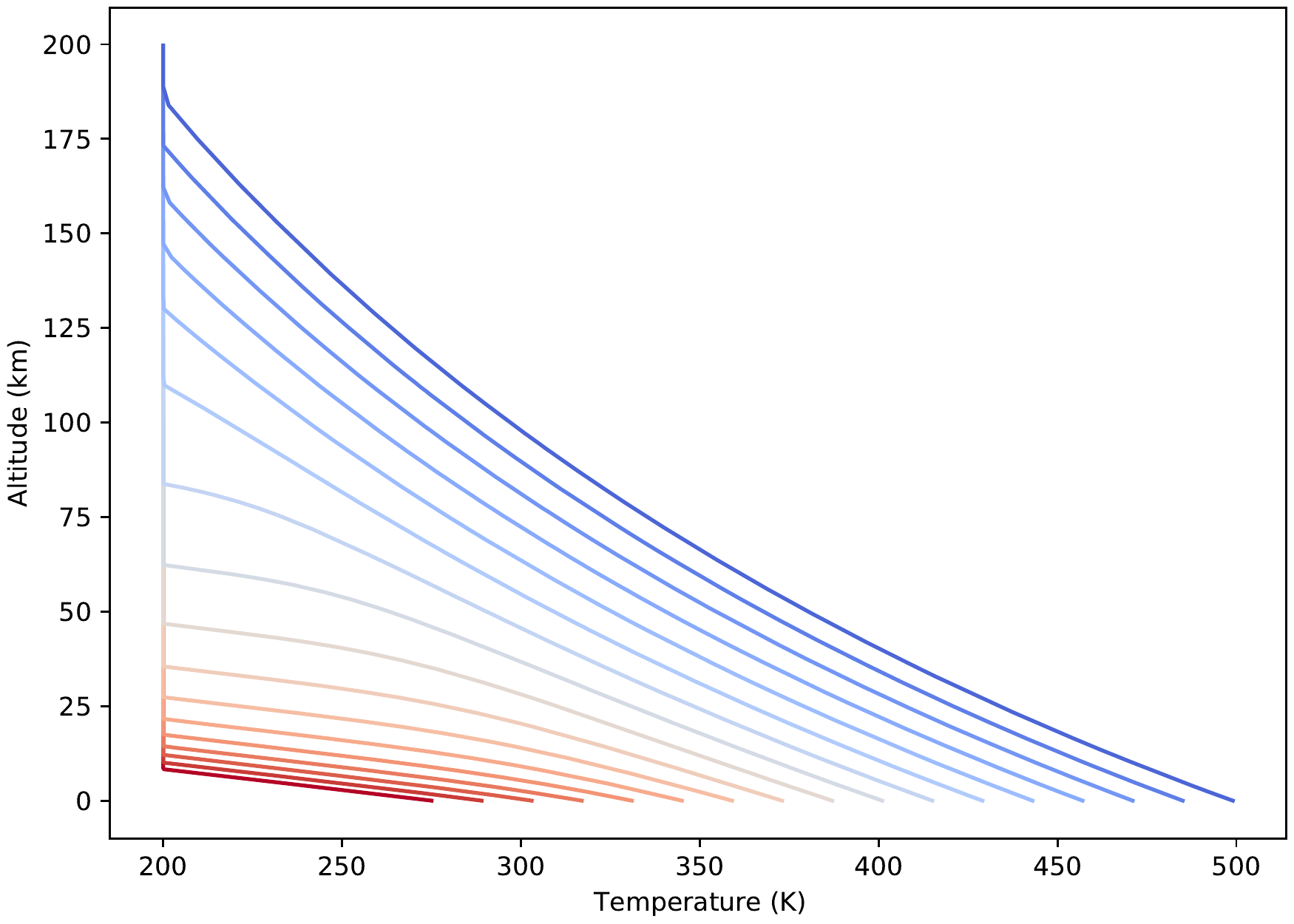}
    \caption{Temperature profiles computed with the Conv\_K88 scheme for different surface temperatures with P$_{\textrm{N}_2}$=10~bar. This figure highlights the transition from dry to wet steeper profiles.}
    \label{fig_T_profiles}
\end{figure}

\textbf{At intermediate mixing ratios (the OLR overshoot),} the corresponding intermediate temperatures corresponds to the transition between an N$_2$ -dominated and an H$_2$O-dominated atmosphere (\fig{fig_N2_mixing_ratio}).
First, this induces a transition between a foreign broadening and a self-broadening of the water absorption lines. As the self-broadening is stronger than the foreign broadening, the absorption increases strongly during this transition. 
As shown in \fig{N2_broad}, the overshoot itself exists through this broadening transition and may be missed by neglecting the change of broadening species (for more details, see \sect{sub_broad}).

Second, the transition between an N$_2$ -dominated and an H$_2$O-dominated atmosphere induces a transition between a dry and a wet adiabatic lapse rate (\figs{fig_adiabat}{fig_T_profiles}). 
Even if N$_2$ is still dominant (volume-mixing ratio below 0.2), the water pressure becomes non-negligible (\figs{fig_N2_mixing_ratio}{fig_OLRvsvmr}). 
Because of that, a small increase in the surface temperature tends to turn the dry adiabatic lapse rate into a wet adiabatic lapse rate. As the adiabat becomes wetter, the structure of the atmosphere changes and the temperature decreases more slowly with the altitude (\fig{fig_T_profiles}), which tends to allow more water in the upper layers.
This has the effect of increasing the size of the atmosphere strongly and quickly (\fig{fig_N2_mixing_ratio}).
Due to this positive feedback effect, the transition of dominant species can be done over a small temperature range (\fig{fig_T_profiles}). 

Nevertheless, we can see that the transition between wet and dry atmospheres spans a wider temperature range for higher nitrogen pressures. If N$_2$ pressure is higher, H$_2$O pressure required to be vapour dominated is higher, and thus the required surface temperature is higher. 
Therefore, the temperature difference between the first visible effects of non-negligible vapour quantity and a vapour-dominated atmosphere is bigger. That is the reason why the overshoot spans a wider temperature range for 10~bar of nitrogen (\fig{fig_PyRADS-Conv1D}).

Finally, the transition of dominant species also induces a transition of the mean molecular weight, from nitrogen molecular weight to water molecular weight. 
As explained in detail in \sect{sub_mol_weight}, changing the mean molecular weight affects both the atmospheric profile and the radiative transfer calculation. Increasing molecular weight of the background gas tends to reduce the $R/c_p$ value, and thus the OLR decreases. On the other hand, increasing the molecular weight of the background gas also tends to reduce the opacity of the atmosphere, and thus the OLR increases. 
As shown in \fig{fig_mol_weight}, this second effect is stronger; therefore, as shown on the bottom panel of the \fig{fig_mol_weight}, it increasing the molecular weight of the background gas means increasing the OLR. 
When water becomes dominant, the mean molecular weight - and all the physical properties of the atmosphere - tend towards water values, and the OLR converges on the Simpson-Nakajima limit.
Consequently, the height of the overshoot is mainly determined by the molecular weight of the background gas (nitrogen here), while the overshoot itself is due to the broadening transition. 

\subsection{Physical effects of the nitrogen}
In this section, we discuss the two main effects of the nitrogen in detail. These are introduced in \sect{sub_overshoot}. The first one is the broadening effect (\sect{sub_broad}), and the second one is the mean molecular weight effect (\sect{sub_mol_weight}).
\subsubsection{Broadening effect}
\label{sub_broad}

In radiative transfer calculations, whatever the method used, the half width at half maximum (HWHM) of the absorption lines due to the pressure broadening can be described by Eq.~\ref{eq_gamma} from \cite{gordon_hitran2016_2017-1}:
\begin{multline}
\label{eq_gamma}
\gamma(P_{tot},P_{\mathrm{H_2O}},T)=\left(\frac{T_{ref}}{T}\right)^{n_{\mathrm{N_2}}}\gamma_{\mathrm{N_2}}(P_{ref},T_{ref})~(P_{tot}-P_{\mathrm{H_2O}}) \\
+\left(\frac{T_{ref}}{T}\right)^{n_{\mathrm{H_2O}}}\gamma_{\mathrm{H_2O}}(P_{ref},T_{ref})~P_{\mathrm{H_2O}},
\end{multline}
where P$_{tot}$ is the total pressure, P$_{\mathrm{H_2O}}$ the partial water pressure, and T the temperature. $\gamma_{\mathrm{N_2}}(P_{ref},T_{ref})$ and $\gamma_{\mathrm{H_2O}}(P_{ref},T_{ref})$ are the HWHM of the foreign- and the self-pressure broadening at the reference temperature. 
As explained previously, we assume that the nitrogen temperature dependency exponent is equal to the air exponent: $n_{\mathrm{N_2}} = n_{\mathrm{AIR}}$. This assumption is motivated by the lack of an experimental value for N$_2$ alone, but the difference should be small. For the same reasons, we also assume that the self-component of the pressure-induced shift of the line centre ($\delta$ in \citealt{gordon_hitran2016_2017-1}) is equal to that of the air component: $\delta_{self}=\delta_{AIR}$.

\begin{figure}[h!]
    \centering\includegraphics[width=\linewidth]{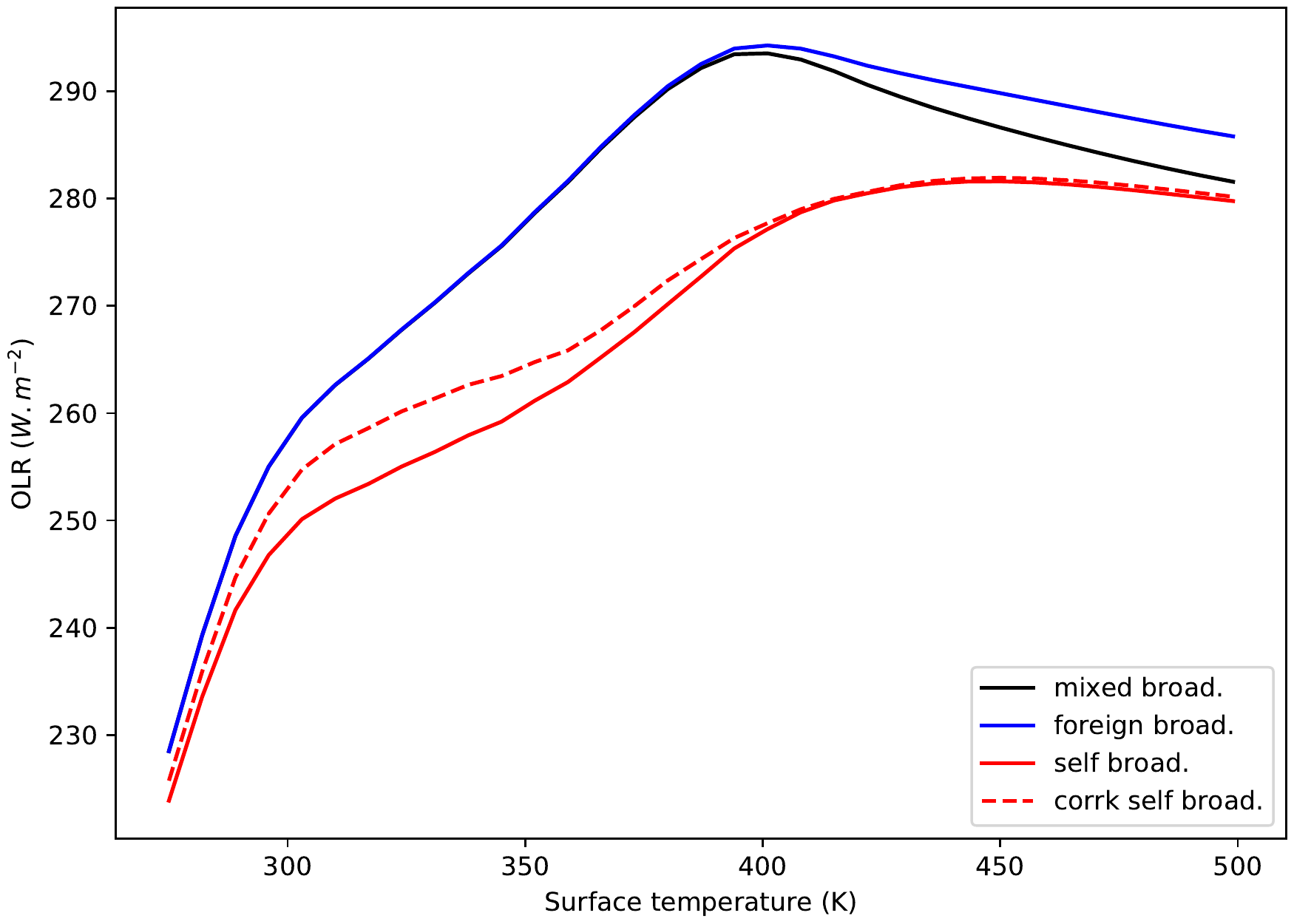}
    \caption{OLR as a function of the temperature for P$_{\textrm{N}_2}$=10 bar. The full lines were computed with PyRADS-Conv1D. The full red line corresponds to the OLR curve for a self-broadening only, and the full blue line corresponds to a foreign broadening. The full black line corresponds to a mixed broadening, such as that presented in the top panel of \fig{fig_PyRADS-Conv1D}. The dotted red line is obtained using \texttt{Exo\_k} with a correlated-k table made with pure water absorption lines (i.e. self-broadening only). Here, both self- and foreign continua are included in the computations.}
    \label{N2_broad}
\end{figure}

If the background gas is a trace gas, it is possible to neglect the HWHM of the foreign broadening and assume that the line broadening only depends on the self-broadening (\eq{eq_gamma_self}). 
However, as explained before, for low surface temperatures the water pressure is limited by the pseudo-adiabatic hypothesis, while the nitrogen pressure is not. Therefore, by only considering the self-broadening to compute the pressure broadening (\eq{eq_gamma_self}), the HWHM and thus the absorption is largely over-estimated, which leads to incorrect OLR values where the overshoot is hardly visible (red lines in \fig{N2_broad}):
\begin{multline}
\label{eq_gamma_self}
\gamma_{\rm self}(P_{tot},T) = \left(\frac{T_{ref}}{T}\right)^{n_{\mathrm{H_2O}}}\gamma_{\mathrm{H_2O}}(P_{ref},T_{ref})~P_{tot},
\end{multline}
where P$_{tot}$ is the total pressure. This effect is strong in our case study because of the transition between an N$_2$ -dominated and an H$_2$O-dominated atmosphere. The black curve in \fig{N2_broad} highlights the transition between the N$_2$-dominant (i.e. foreign) and the H$_2$O-dominant (i.e. self) broadening, corresponding to the curves proposed in \fig{fig_PyRADS-Conv1D}.

It is possible to build a correlated-k table from high-resolution spectra of pure water even by neglecting the foreign broadening. This possibility is commonly used to skip the tedious step of calculating a large number of spectra in the case of a unusual mix of gases. 
By doing that, we assume a self-broadening only, and thus we over-estimate the absorption (dotted red line in \fig{N2_broad}) as explained just above.
Regarding our conclusions, the error induced by this method is only negligible if the background gas is a trace gas. However, \cite{amundsen_treatment_2017} discussed different possibilities to easily produce correlated-k for mixed gases. 
They show that in particular cases, correlated-k table built by mixing the tables of pure species, that is by neglecting the mutual broadening, may produce accurate results with a greater flexibility than by creating pre-mixed correlated-k tables.

It can be counter-intuitive to state that an OLR curve produces an overshoot by only assuming the foreign broadening but not by only assuming the self-broadening (\fig{N2_broad}). This may be explained by a subtle effect of the broadening power of water and nitrogen. 
As the self-broadening is stronger than the foreign one, the absorption lines of the `self-only case' are probably already more broadened than for the `foreign-only case'. 
Therefore, the overshoot effect described previously is probably weaker if we only consider the self-broadening. Nevertheless, a deeper study of this point is needed before any conclusion can be drawn.

\subsubsection{Molecular weight effect}
\label{sub_mol_weight}

As discussed in \cite{pierrehumbert_principles_2010}, the addition of a background gas modifies the atmospheric structure, which plays an important role on the shape of the OLR curve. We can study this effect by analysing the effect of the molecular weight of the background gas on the atmospheric profile and on the radiative transfer calculation separately.
For convenience and as we only analysed tendencies, we chose to use the Conv\_D16 scheme and the \texttt{Exo\_k}-RT radiative transfer method. In the same way, radiative properties of the background gas (mainly the broadening) are those of nitrogen in order to only study the effect of the molecular weight on the OLR value.\\

\begin{table}[ht]\footnotesize\centering
\setlength{\doublerulesep}{\arrayrulewidth}
\captionsetup{justification=justified}
\caption{Molecular weight ($M$), heat capacity ($c_p$), and molecular heat capacity ($c_{p,molec}$) for different background gases (Gas), where $c_{p,molec}=c_p\times M$. The heat capacity values are taken at $0^{\circ}$C and 1\,bar from \cite{pierrehumbert_principles_2010}.}
\label{table_Mn}
\begin{tabular}[c]{cccc}
    \hline\hline\hline
    Gas  & M (g.mol$^{-1}$) & $c_{p}$ (J.g$^{-1}$.K$^{-1}$) & $c_{p,molec}$(J.mol$^{-1}$.K$^{-1}$)  \\
    \hline
    He & 4.003 & 5.196 & 20.80 \\
    H$_2$ & 2.016 & 14.23 & 28.69 \\
    N$_2$ & 28.02 & 1.037 & 28.02 \\
    O$_2$ & 31.99 & 0.916 & 29.30  \\
    CO$_2$ & 44.01 & 0.820 & 36.09  \\
    \hline\hline\hline
\end{tabular}\normalsize
\end{table}

\begin{figure}[h!]
    \centering\includegraphics[width=\linewidth]{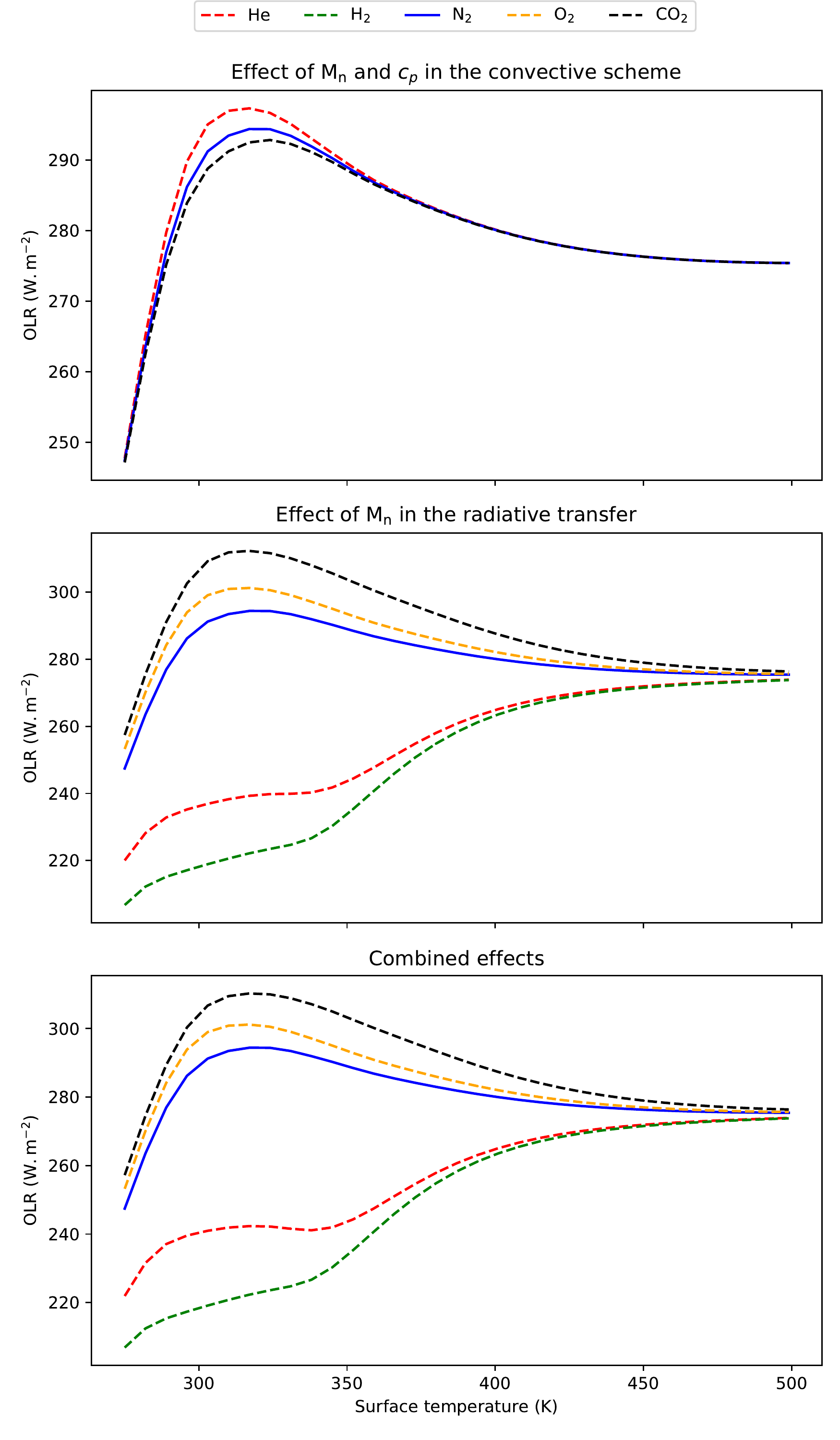}
    \caption{OLR as a function of the surface temperature by only using the physical properties of other gases in the convective scheme calculation (top panel), in the radiative transfer calculation only (middle panel) and in both of them (bottom panel). The full blue line corresponds to the curve presented in the top panel of \fig{fig_PyRADS-Conv1D} with the true value of the nitrogen molecular weight. The dotted lines correspond to $M_n$ and $c_p$ values of other gases (see \tab{table_Mn}). Here, we used the Conv\_D16 scheme and the \texttt{Exo\_k}-RT radiative transfer method assuming 1\,bar of background gas.}
    \label{fig_mol_weight}
\end{figure}

\textbf{With regard to the effect on the atmospheric profile,} by rewriting the equation proposed by \cite{ding_convection_2016} (see \eq{eq_pierrehumbert16}) and assuming a constant number of background molecules ($n_n$), it is possible to show that the adiabatic lapse rate depends on the molecular weight of the background gas ($M_n$) only through the $R_n/c_{pn}$ value (in red in \eq{eq_Ding_Mn}), where $R_n$ is the specific gas constant of the background gas (\eq{eq_Ding_Mn}): 
\begin{equation}
        \label{eq_Ding_Mn}
        \frac{d\ln P}{d\ln T}=\frac{P_c}{P}\frac{L}{R_cT}+\frac{P_n}{P}\times\frac{\color{red}\frac{c_{pn}}{R_n}\color{black}+\left(c_{pc}+\left(\frac{L}{R_cT}-1\right)\frac{L}{T}\frac{m_c}{R^* n_n}\right)\alpha_c}{1+\frac{L}{T}\frac{m_c}{R^* n_n}\alpha_c}
.\end{equation}
Indices of variables in \eq{eq_Ding_Mn} indicate what gas is considered: $c$ for the condensable gas (water here) and $n$ for the background gas. Here, $P$ is the pressure, $T$ the temperature, $M$ the molecular weight, $R^*$ the perfect gas constant, $R$ the specific gas constant, $L$ the latent heat, and $c_p$ the heat capacity.
By changing the background gas, both $M_n$ and $c_{pn}$ change as shown in \tab{table_Mn}. However, for diatomic gases (N$_2$, O$_2$, H$_2$ ...), the molecular heat capacity ($c_{p,molec}=c_p\times M$) is remarkably similar. By definition, the $R_n/c_{pn}$ value is also remarkably similar whatever the considered diatomic gas.
This means that for such gases, most of the atmospheric profiles will be the same, as will the OLR curve as shown on the top panel of \fig{fig_mol_weight}.
For triatomic molecules such as CO$_2$, the $R_n/c_{pn}$ value decreases, which means that the temperature decreases more slowly with altitude.
Therefore, the atmosphere warms up and because of the pseudo-adiabatic hypothesis, the total amount of water increases, and the OLR decreases. \\

\textbf{Concerning the effect on the radiative transfer,} the molecular weight of the background gas has a much stronger impact on the radiative transfer calculation. By assuming a radiatively inactive background gas, the opacity of the atmosphere ($\tau$) can be simply written by \eq{eq_tau_M}:
\begin{equation}
    \label{eq_tau_M}
    \tau=\int \frac{x_c \sigma_c}{x_c M_c +x_n M_n} \frac{d p}{N_A g}
,\end{equation}
where $x_i$, $\sigma_i$ are the volume-mixing ratio and the cross-section per molecule (in $m^2/molecule$) of the species $i$. $N_A$ is the Avogadro number. 

Interestingly, if the the molecular weight of the background gas ($M_n$) changes without altering the atmospheric temperature profile (i.e. if $R_n/c_{pn}$ remains constant), then the vapour-volume-mixing ratio profile does not change either because it only depends on $T(p)$. In other words, at any pressure level, the water-molecule-to-background-gas ratio stays the same. However, rather counter intuitively, changing $M_n$ at a constant surface pressure changes the total number of molecules of background gas, and hence the total number of water molecules, although the ratio and total mass stay the same. 

Therefore, for a given atmospheric profile, if the molecular weight of the background gas ($M_n$) decreases, the mass of water vapour  in the atmosphere increases and the latter becomes more opaque, as demonstrated by \eq{eq_tau_M}. This leads to a decrease in the OLR, as shown on the middle panel of \fig{fig_mol_weight}.\\ 

\textbf{Concerning the global effect of the molecular weight,} as is visible in \fig{fig_mol_weight}, the effect of the molecular weight on the radiative transfer calculation (middle panel) is largely dominant compared to the modification of the atmospheric profile (top panel). 
For this reason, the OLR curve obtained by taking into account both effects (bottom panel) is similar to the one for which we take into account only the effect on the radiative transfer.

For gas mixtures with other inactive gases (e.g. O$_2$, H$_2$), the proposed analysis may help to understand how the OLR evolves regarding changes of the mean molecular weight of the atmosphere. For example, gases lighter than N$_2$ - such as H$_2$ - tend to reduce the OLR, as shown by previous studies \citep[e.g.][]{koll_hot_2019}. For radiatively active gases such as CO$_2$, the molecular weight effect still plays a role but may become a second-order process compared to the absorption of the gas itself (see e.g. \sect{co2}). However, when changing the mixture, the radiative properties of the atmosphere will change and the OLR will be slightly impacted.

It is interesting to notice that when the global effect of the molecular weight is taken into account (Bottom panel on \fig{fig_mol_weight}), there is a qualitative change in the shape of the OLR curve, and even a hump when using the He gas properties.
This hump, which arises below the Simpson–Nakajima limit, is discussed in \cite{koll_hot_2019} as a 'souffl\'e' effect with the example of H$_2$. This effect is more pronounced in \cite{koll_hot_2019} because they assume a grey gas for the radiative transfer calculation. 
On the He curve (\fig{fig_mol_weight}), the two physical effects described in \sect{sub_overshoot} are clearly visible. First, the overshoot itself, due to the transition between a foreign-broadening and a self-broadening (\sect{sub_broad}) and, second, the convergence of the physical properties of the atmosphere toward pure steam atmosphere properties (e.g. the mean molecular weight).

In this section, we show that the height of the overshoot is given by the background gas molecular weight, but \fig{fig_PyRADS-Conv1D} shows that the OLR overshoot assuming 0.1~bar of nitrogen is weaker than it is for higher nitrogen pressures.
As presented previously, the transition between nitrogen-dominated and vapour-dominated atmospheres is key to understanding the overshoot, but this induces a minimal N$_2$ pressure to achieve this transition. With 0.1~bar of nitrogen, low vapour pressures induced by low temperatures are sufficient to be non-negligible, with a volume-mixing ratio higher than $5\times10^{-5}$ (see \figs{fig_N2_mixing_ratio}{fig_OLRvsvmr}). Consequently, the atmosphere is never fully nitrogen dominated, and the height of the OLR overshoot is weaker. \cite{koll_hot_2019} defined a similar \textit{\emph{'dilute limit'}}.

\section{Discussion}
\label{sec_discussion}

In this section, we discuss our results (\sect{sec_disc_results}), and we propose explanations to understand the differences in the results of the literature (\sect{sec_disc_literature}).

\subsection{Discussion of the results}
\label{sec_disc_results}
The OLR curves presented in \fig{fig_PyRADS-Conv1D} are proposed as reference curves because they include all the major physical processes of an H$_2$O+N$_2$ atmosphere. However, these processes were tested for Earth-like planets with surface temperatures between 275~K and 500~K.
We are confident in the applicability of our conclusions with other similar inactive background gases (e.g. H$_2$), particularly concerning the hierarchy of importance induced by the sensitivity tests presented in \tab{table_tests}. 
In the same way, \sects{sub_broad}{sub_mol_weight} describe the effect of the broadening and of the molecular weight with the example of nitrogen, but our conclusions are applicable for other background gases.

This work can be a first step towards a complete overview of effects of different background gases on an Earth-like atmosphere.
The accuracy obtained on OLR values in this paper is higher than the precision required for a GCM simulation regarding to other uncertainty sources, but presented sensitivity tests could be useful for GCM intercomparisons \citep[e.g.][]{yang_differences_2016, yang_simulations_2019, fauchez_trappist_2021}.
Also, we do not discuss the variations of the absorbed solar radiation (ASR) in this work (which is necessary to compute up to the top-of-atmosphere radiative budget), but it is a major quantity to study the runaway greenhouse and compute the inner edge of the HZ.

In our model, we assume a relative humidity (RH) equal to unity, but this is a strong assumption that can lead to incorrect estimations of the OLR, particularly regarding values from most complex and accurate simulations using GCMs \citep[e.g.][]{leconte_increased_2013}. A good improvement should be to adapt the RH of our model as prescribed by \cite{leconte_increased_2013}, or to use a variable value such as that of \cite{goldblatt_low_2013}.
\cite{ramirez_can_2014} used 1D self-consistent RH parametrisation, which can also be a solution to improve our calculation.
In the same way, we followed the description
of the stratosphere from \cite{kasting_runaway_1988}, assuming a constant temperature and mixing ratio. This is valid when the stratospheric part of the atmosphere is small compared to the troposphere, that is at a high temperature. At a low temperature, when the adiabat is close to a dry adiabat, this assumption should induce strong inaccuracies. 
It could be interesting to quantify the differences between our radiative-convective model with a time-marching model that takes into account the radiative heating of the star. This may be helpful to understand the error induced by the stratospheric hypothesis.
We parametrise PyRADS-Conv1D to reach an error at most equal to 1W/m$^2$, but we neglect the dynamics and assume a cloud free atmosphere. This assumption is motivated by the fact that the dynamical effects (e.g. advection) are at best parametrised in 1D models. 
However, the strong greenhouse power of the clouds and the redistribution of the heat and humidity due to advection have a strong impact on the OLR, as shown by many studies \citep[e.g.][]{yang_stabilizing_2013,leconte_increased_2013,yang_simulations_2019}.

\app{sub_vertical_grid} discusses the definition of the vertical grid and the minimal required number of atmospheric levels. This can be a critical point for GCMs where the number of levels is limited. 
The height of the atmosphere strongly increases with increasing surface temperature, that is the amount of vapour. This can become problematic if the number of levels is not sufficient - or if the atmosphere is not extended enough - leading to a truncation of the 'radiatively active part' of the atmosphere. 
A solution would be to modify the vertical grid to keep a low vapour pressure in the upper layers and to adapt the altitude of the atmospheric levels to accurately represent the lower part of the atmosphere.

\subsection{Literature comparison}
\label{sec_disc_literature}
A large model intercomparison was done in \cite{yang_differences_2016} for an H$_2$O+N$_2$ atmosphere with 376ppmv of CO$_2$. They highlighted the fact that differences between the models increase with the temperature and reach 25~W/m$^2$ at 360~K. 
This is similar to the range of values we observed in \fig{model_comparison} at low nitrogen pressure with our set of models. Thanks to our multiple sensitivity tests, we are able to explain this wide range of results by discussing parametrisations and assumptions used in different 1D models from the literature.

\cite{kopparapu_habitable_2014} added 350~ppm of CO$_2$ to the radiative transfer calculation (Ravi K. Kopparapu, personal communication) compared to other results presented in \fig{model_comparison}. This can strongly modify the OLR, as explained in \app{co2}. However, the computations done by that group show that even with a pure N2 atmosphere they do not reproduce the overshoot of the OLR (anonymous referee, personal communication).
They used the adiabat proposed by \cite{kasting_runaway_1988} to compute the atmospheric profiles, assuming 101 levels and a constant stratospheric temperature equal to 200~K \citep{kopparapu_habitable_2013}. The radiative transfer calculation is based on a correlated-k method that uses a combination of the HITRAN and HITEMP databases \citep{kopparapu_habitable_2013}.
To compute the water continuum, \cite{kopparapu_habitable_2014} used BPS \citep{paynter_assessment_2011}, while we used MT\_CKD in PyRADS-Conv1D. As presented in \app{sub_continuum}, the choice of the continuum may shift the asymptotic limit of the runaway greenhouse.
Finally, \cite{kopparapu_habitable_2014} did not obtain an OLR overshoot, unlike other studies presented in \fig{model_comparison}. 
They used separated H$_2$O and N$_2$ correlated-k coefficients instead of mixed coefficients (anonymous referee, personal communication), and as highlighted by our sensitivity tests, the easiest way to miss the overshoot is to neglect the foreign broadening of the water absorption line. Therefore, this assumption could explain a part of the difference observed in \fig{model_comparison}.

To compute the radiative transfer of the atmosphere, \cite{goldblatt_low_2013} used a line-by-line code called SMART \citep{meadows_ground-based_1996} between 0 and 30000 cm$^{-1}$ (for the water) using HITEMP2010. The HITRAN and HITEMP databases are not strictly equal and may produce different OLR values \citep[e.g.][]{goldblatt_low_2013}, but this is negligible below 350K \citep{kopparapu_habitable_2013}. 
They also compute their own water continuum which slightly underestimates the absorption \citep{goldblatt_low_2013}. Moreover, as shown in \fig{fig_mtckd}, different continua may induce a different estimation of the Simpson-Nakajima limit.
This may partially explain the high OLR values obtained by \cite{goldblatt_low_2013}.
They also assume a variable RH, which is more accurate than assuming an RH equal to unity, as is frequently done in 1D radiative-convective models (including PyRADS-Conv1D). As shown by \cite{leconte_increased_2013}, this may have a strong impact on the shape of the OLR curve and also on the Simpson-Nakajima asymptotic value. This is probably the main difference between the results of \cite{goldblatt_low_2013} and others in \fig{fig_mtckd}.

The asymptotic value of the runaway greenhouse obtained by \citet{zhang_how_2020} is much higher than that from \citet{goldblatt_low_2013}, \cite{kopparapu_habitable_2014}, or ours. They used ExoRT\footnote{\url{https://github.com/storyofthewolf/ExoRT}}, which is derived from the GCM named ExoCAM. GCMs sometimes contain strong assumptions that are necessary regarding the computation time; thus, additional analyses are required to understand the difference between the results of \cite{zhang_how_2020} and those of others.

In the same way, the 1D reverse version of the LMD-Generic model (kcm1d; see \citealt{turbet_runaway_2019}) provides results that are shifted lower compared to PyRADS-Conv1D, \texttt{Exo\_k,} and PyRADS. This is due to both the two-stream solution used (constrained by the LMD-Generic requirements) and the correlated-k interpolation scheme. In kcm1d, the average angle of the outgoing flux used ($\mu_0$) is fixed by the Hemispheric mean method.

\section{Conclusion}
\label{sec_conclusion}
In this work, we built a new 1D radiative-convective model named PyRADS-Conv1D -- based on \citet{koll_hot_2019} and \citet{marcq_thermal_2017} -- to propose OLR reference curves (\fig{fig_PyRADS-Conv1D}) for an H$_2$O+N$_2$ atmosphere. Through multiple sensitivity tests (see \app{sub_sensitivity}), we were able to identify the most important physical processes required to accurately model such atmospheres. These reference curves confirm the occurrence of an overshoot of the OLR relatively to the Simpson-Nakajima limit by adding nitrogen in a pure vapour atmosphere. We provide an accurate H$_2$O+N$_2$ correlated-k table that reproduces the results obtained by the line-by-line calculation.

We show that the overshoot is due to a non-usual transition between an N$_2$-dominated atmosphere and an H$_2$O-dominated atmosphere (see \sect{sec_results}). This transition challenges the modelling by making important, usually second-order, processes.
More precisely, we explain that the OLR overshoot is due, firstly, to a transition between a foreign or self-broadening of the water absorption lines (see \sect{sub_broad}), and, secondly, to a transition between a dry adiabatic lapse rate and a moist adiabatic lapse rate.
In other words, the overshoot itself is due to a broadening transition, and its height is determined by the molecular weight of the background gas (nitrogen here) through a modification of the dry part of the adiabat (see \sect{sub_mol_weight}). A heavier gas induces a stronger overshoot, and a lighter one induces a weaker overshoot.

Our sensitivity tests also allow us to highlight that differences between previous studies are mainly due to missing physics or inaccurate or over-simplified parametrisations. 
For this reason, we list the most important physical processes and parametrisations needed to obtain an accurate value of the OLR for the considered atmosphere, and we quantify the error or the uncertainty induced by each of them (see \tab{table_tests}). 
Several of these physical processes induce negligible errors if they are omitted, but some of them may induce a large error, large enough to lead to their missing the OLR overshoot. 
First of all, the foreign broadening cannot be neglected (\fig{N2_broad}), otherwise the overshoot vanishes because the pressure broadening of the water absorption lines is over-estimated. Secondly, the water continua (H$_2$O-H$_2$O, H$_2$O-N$_2$) should be carefully chosen because they have a non-negligible impact on the computed OLR value, about a few watt per square meter, and on the value of the asymptotic limit of the runaway greenhouse (\app{sub_continuum}).
The convection scheme may also induce a difference in the estimation of the content of water vapour by using the perfect gas approximation.
Finally, the two-stream solution used (i.e. the $\mu_0$ angle chosen, see \app{sub_two-stream}) and the correlated-k interpolation scheme may modify the asymptotic limit of the runaway greenhouse by inducing a OLR difference of several W.m$^2$.

Following a similar approach to \cite{kasting_runaway_1988} or \cite{kopparapu_habitable_2013} to compute the inner edge of the HZ leads us to consider the maximum OLR value as a threshold value for the onset of the runaway greenhouse. The OLR overshoot we obtained tends to validate that the inner limit of the HZ depends on the nitrogen pressure \cite[e.g.][]{ramirez_effect_2020}. 
Moreover, this inner limit is probably closer to the host star compared to previous studies that missed the overshoot \citep[e.g.][]{kopparapu_habitable_2014}.

Modelling an H$_2$O+N$_2$ atmosphere using a GCM may probably provides very different OLR values because of the dynamics and the radiative effects of the clouds \citep[e.g.][]{yang_stabilizing_2013,leconte_increased_2013,wolf_assessing_2017,yang_simulations_2019}.
The radiative effect of the clouds is usually a dominant process in the computation of the OLR \citep[e.g.][]{leconte_increased_2013,turbet_nightside_nodate}, and the water-dominated atmosphere induced by high temperatures is probably highly cloudy. 
\cite{leconte_increased_2013} also showed that the stabilising effect of the Hadley circulation shifts the greenhouse limit higher. For these reasons, we plan to revisit this work using a GCM that includes large-scale dynamics and clouds.
Nevertheless, the processes we describe in this work and the pressure-broadening transition should still play a role in such simulations.  

\section*{Acknowledgements}
This work has been carried out within the framework of the National Centre of Competence in Research PlanetS supported by the Swiss National Science Foundation. The authors acknowledge the financial support of the SNSF. This project has received funding from the European Union’s Horizon 2020 research and innovation program under the Marie Sklodowska-Curie Grant Agreement No. 832738/ESCAPE and under the European Research Council (ERC) grant agreement n$^\circ$679030/WHIPLASH. M.T. thanks the Gruber Foundation for its support to this research.

\bibliographystyle{aa}
\bibliography{biblio}

\clearpage

\appendix

\section{Sensitivity tests}
\label{sub_sensitivity}
A number of sensitivity tests were performed to constrain the relevant different physical processes we have to take into account to compute the OLR of the atmosphere we consider (\app{sub_sensi_study}). We tested also the errors induced by the convergence of several numerical parameters - or by numerical assumptions (\app{sub_num_study}). An overview of these tests is proposed in \tab{table_tests}.

\begin{table*}\footnotesize\centering
\setlength{\doublerulesep}{\arrayrulewidth}
\captionsetup{justification=justified}
\caption{Overview of sensitivity tests performed with the impact on the computed OLR value. The \textit{Original param.} column indicates the parametrisation used in the considered model, and the \textit{Tested param.} column shows the tested parametrisation. The OLR difference between original and modified parametrisations is indicated in the \textit{OLR difference} column.}
\label{table_tests}
\begin{tabular}[c]{cccc}
    \hline\hline\hline
    Model tested  & Original param. & Tested param. & OLR difference  \\
    \hline
    PyRADS-Conv1D & Lorentz shape & Voigt shape & < 1~W.m$^2$ \\
    PyRADS-Conv1D \& \texttt{Exo\_k} & with foreign broad. & without foreign broad. & up to 20~W.m$^2$ \\
    PyRADS-Conv1D & MT\_CKD3.2 [0.1 - 10k]cm$^{-1}$ & MT\_CKD3.2 [0.1 - 20k]cm$^{-1}$ & < 2~W.m$^2$\\
    \texttt{Exo\_k} & MT\_CKD3.2 & MT\_CKD2.5 & < 2W.m$^2$ \\
    \texttt{Exo\_k} & with H$_2$O-N$_2$ cont. & without H$_2$O-N$_2$ cont. & up to 12~W.m$^2$ (at low T) \\
    \texttt{Exo\_k} & with N$_2$-N$_2$ cont. & without N$_2$-N$_2$ cont. & up to 4~W.m$^2$ (at low T) \\
    PyRADS-Conv1D & without water iso. & with water iso. & < 2~W.m$^2$  \\
    kcm1d \& \texttt{Exo\_k} & without CO$_2$ & 376ppm of CO$_2$ &  up to 40~W.m$^2$ \\
    \texttt{Exo\_k} & two-stream: $\bar{\mu}$=0.6 & two-stream: Hemis. mean & $\sim$ 6~W.m$^2$ \\
    PyRADS-Conv1D & Conv\_K88 & Conv\_D16 & < 5~W.m$^2$\\
    \hline
    PyRADS-Conv1D & high res. vertical grid & low res. vertical grid & up to 10~W.m$^2$ \\
    PyRADS-Conv1D & spec. res. 10$^{-2}$ cm$^{-1}$ & spec. res. 10$^{-3}$ cm$^{-1}$ & < 1~W.m$^2$ \\
    \texttt{Exo\_k} & \texttt{Exo\_k} interp. scheme & kcm1d interp. scheme & $\sim$ 3~W.m$^2$\\
    \hline\hline\hline
\end{tabular}\normalsize
\end{table*}

\subsection{Sensitivity studies on physical processes}
\label{sub_sensi_study}
The first set of sensitivity tests is focused on physical processes or hypotheses.
\subsubsection{Shape of the absorption lines}
\begin{figure}[h!]
    \centering\includegraphics[width=\linewidth]{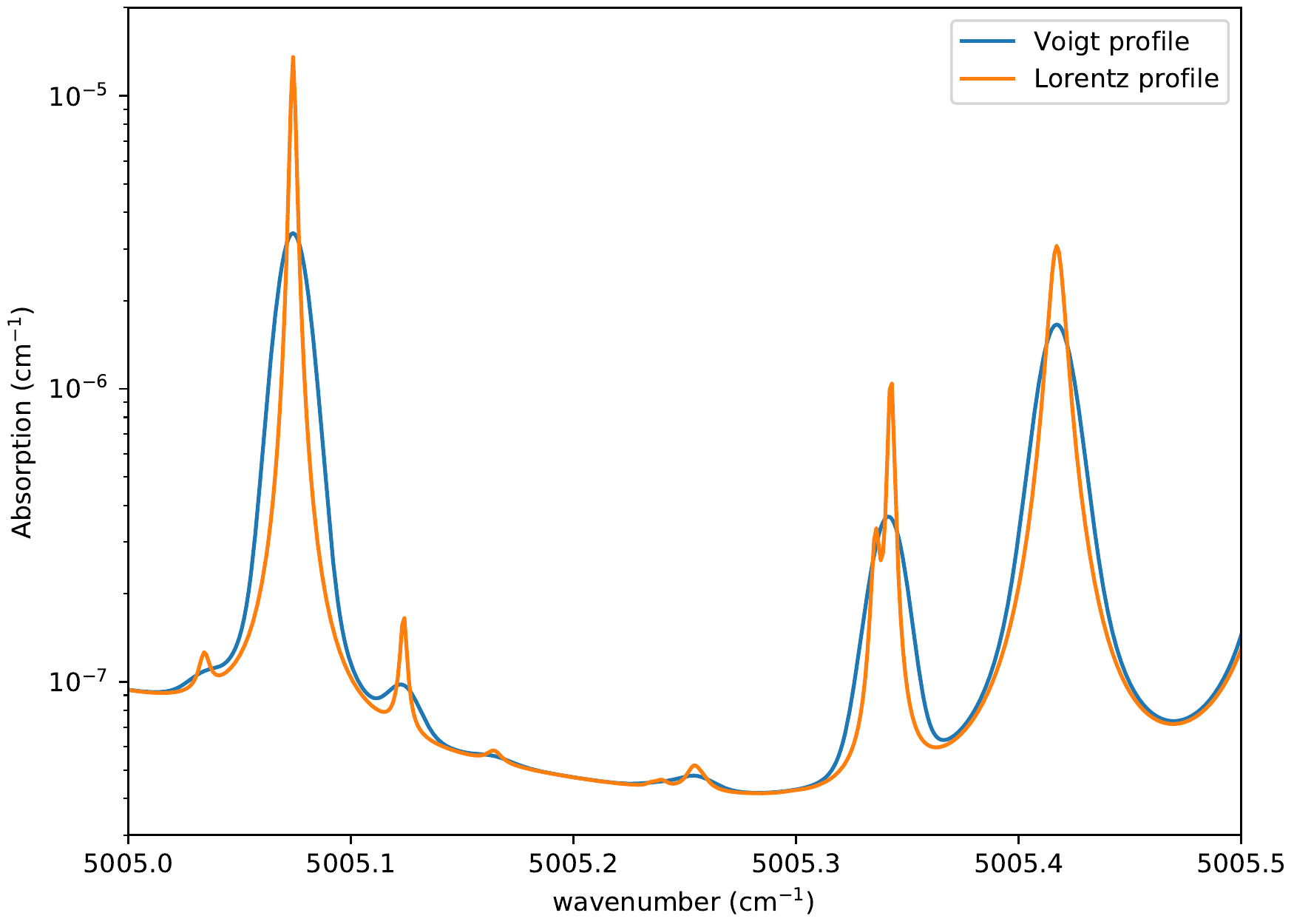}
    \caption{Absorption spectrum assuming a Voigt or a Lorentz profile at the same resolution (0.001~cm$^{-1}$) with P=1000~Pa and T=300~K. As the wings of the absorption lines do not follow a Voigt profile, we only considered the centre of the lines up to 25\,cm$^{-1}$here.}
    \label{fig_voigt_lorentz}
\end{figure}

The most accurate approximation to represent the absorption lines is to assume a Voigt profile, which is a convolution of a Gauss and a Lorentz profile. Unfortunately, there is no analytic solution to compute this profile, and the numerical computation is time expensive.
As explained by \cite{koll_earths_2018}, assuming a Lorentz shape is sufficient (in our case study) to obtain an accurate value of the OLR (Error lower than 1~W.m$^{-2}$).
To create the correlated-k table used in this work, we computed high-resolution spectra assuming a Voigt profile. Being a statistical description of the absorption over a large range of temperature (pressure and mixing ratio values), this method requires extremely accurate initial spectra. 
As shown in \fig{fig_voigt_lorentz}, the Lorentz profile tends to reduce the absorption at the centre of the line and increase it in the wings, but this difference is negligible for the OLR calculation.
At higher pressure, that is higher broadening, the width of the lines is larger and the difference between the Lorentz and the Voigt profile decreases.

\subsubsection{Continua}
\label{sub_continuum}

\begin{figure}[h!]
    \centering\includegraphics[width=\linewidth]{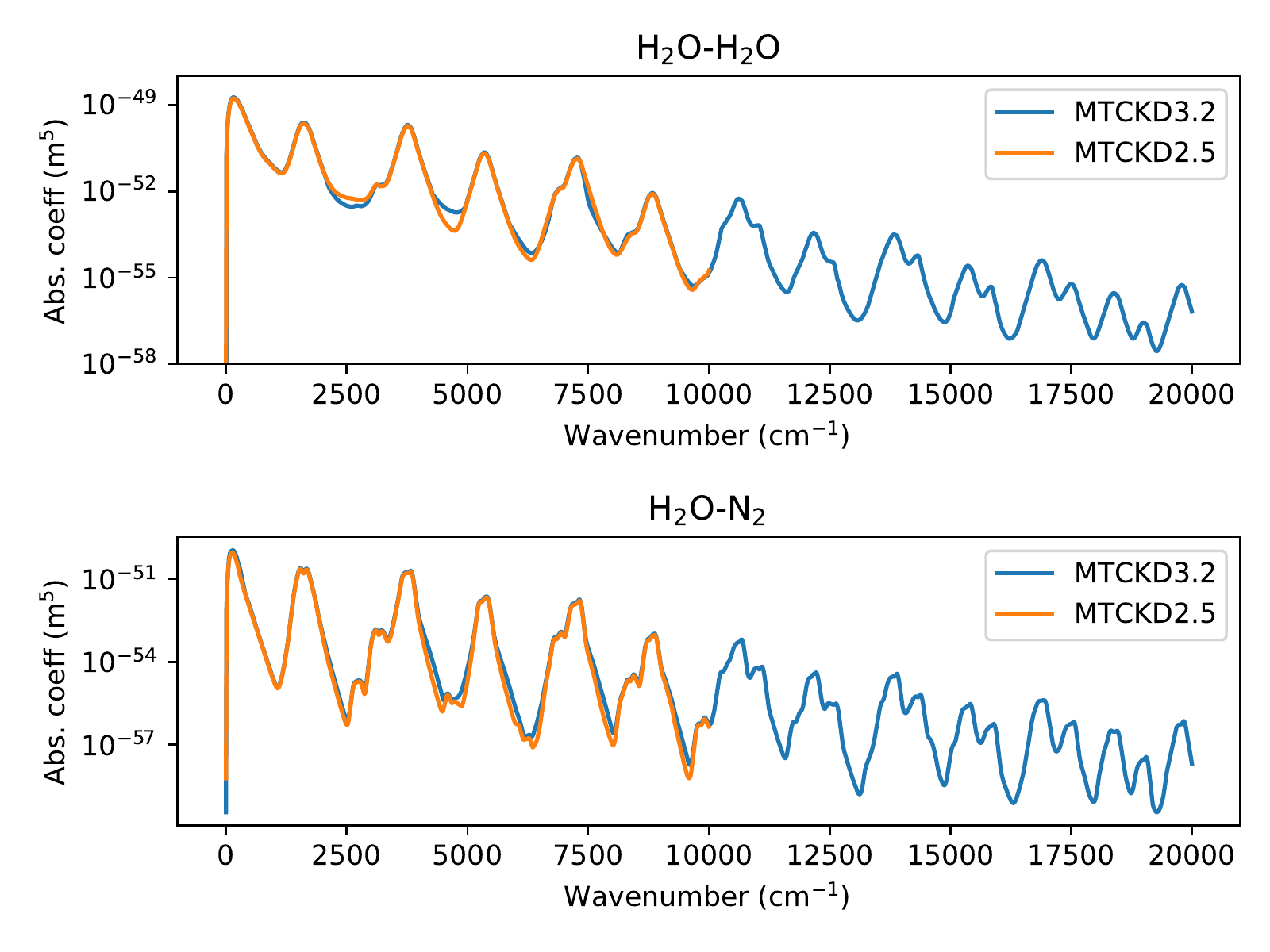}
    \caption{Comparison of H$_2$O-H$_2$O and H$_2$O-N$_2$ continua from MT\_CKD2.5 and 3.2 databases. Here, the y-axis is in logscale.}
    \label{fig_mtckd}
\end{figure}

\begin{figure}[h!]
    \centering\includegraphics[width=\linewidth]{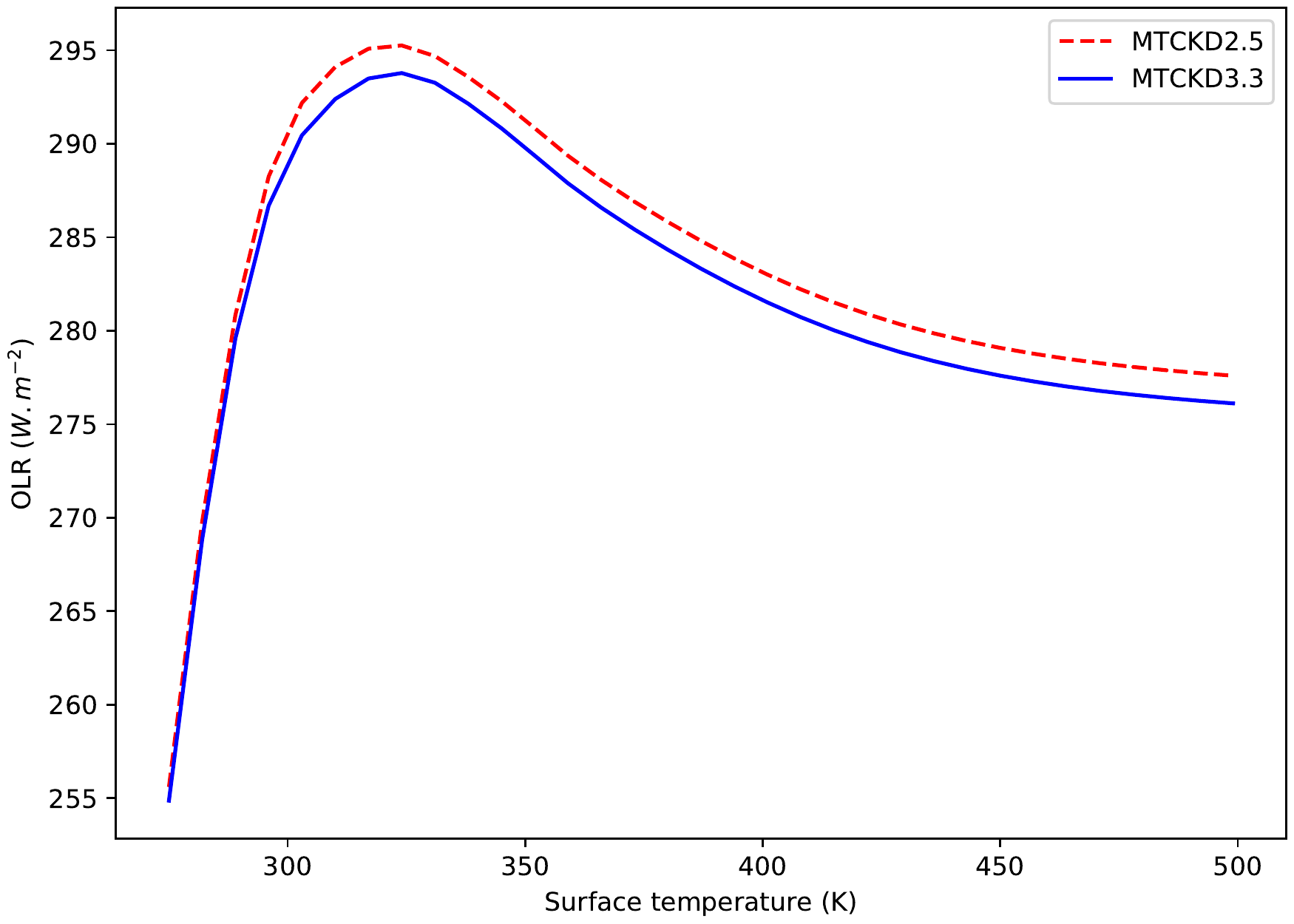}
    \caption{OLR values as a function of the surface temperature for P$_{\textrm{N}_2}$=1~bar for different versions of the MT\_CKD database.}
    \label{fig_mtckd_olr}
\end{figure}

In this work, for both radiative transfer methods, we used the MT\_CKD formalism with a cut-off of the centre of the water absorption lines at 25~cm$^{-1}$ and a removed plinth. A continuum is added to take into account the far-wing absorption.
\cite{yang_differences_2016} showed through a model inter-comparison that a source of difference of the OLR values comes from the radiative transfer, and more particularly from the continua used. 
It is interesting to note that a part of the difference between our results and those of \cite{kopparapu_habitable_2013} is due to the choice of the H$_2$O continuum database. They used BPS \citep{paynter_assessment_2011}, while we used MT\_CKD. 

A spectral comparison of the water continua used in this work is provided in \fig{fig_mtckd}. 
The main difference between MT\_CKD2.5 and MT\_CKD3.2 comes from the wavelength definition domain. The MT\_CKD2.5 continuum is defined between 0~cm$^{-1}$ and 10000~cm$^{-1}$ , while the MT\_CKD3.2 continuum is defined up to 20000~cm$^{-1}$ , as shown in \fig{fig_mtckd}.
Nevertheless, as the vast majority of the absorption of the water is between 0~cm$^{-1}$ and 5000~cm$^{-1}$ , the difference in the OLR value is at most 1.5~W.m$^2$ (\fig{fig_mtckd_olr}). 
We used the MT\_CKD3.2 continua in PyRADS-Conv1D to increase our accuracy but also for computational reasons. 
If the continnum is not defined for a spectral range, it becomes necessary to compute the absorption lines up to a few hundreds of cm$^{-1}$ to accurately represent the spectrum. This increases the computation time drastically.  
The N$_2$-N$_2$ continuum from the HITRAN CIA database even has a negligible impact on the OLR (less than 1~W/m$^2$) at a low temperature for $\mathrm{P_{N_2}}$=10~bar when the nitrogen pressure is largely dominating the vapour pressure.

\subsubsection{Water isotopes}
\label{sub_isotopes}

\begin{figure}[b!]
    \centering\includegraphics[width=\linewidth]{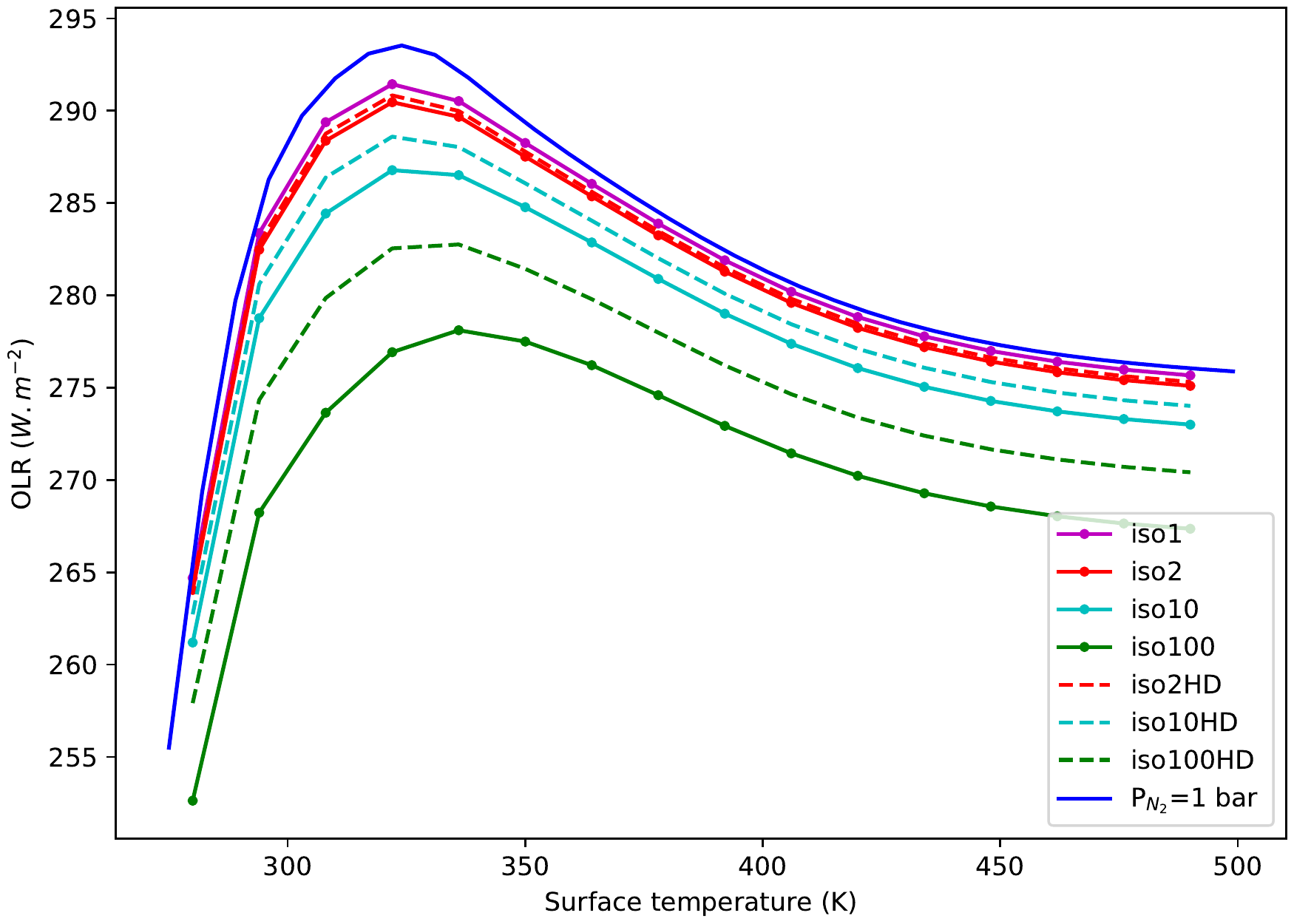}
    \caption{OLR values as a function of the surface temperature for P$_{\textrm{N}_2}$=1~bar for different water isotopes quantities and using PyRADS-Conv1D at high spectral resolution (0.001~cm$^{-1}$).}
    \label{fig_isotopes}
\end{figure}

There is no real consensus to include (or not include) the absorption lines of the secondary water isotopes in the radiative transfer computation. Similarly, there is no consensus as to what their abundances may be in other worlds. To have an idea of the potential impact of water isotopes on the OLR computation for an Earth-like planet, we considered different case studies with different isotopes abundances: Iso1 (terrestrial abundances, \citealt{de_bievre_isotopic_1984,gordon_hitran2016_2017-1}), 
 Iso2 (all the secondary isotopes are two times more abundant),
 Iso10 (all the secondary isotopes are ten times more abundant),
 Iso100 (all the secondary isotopes are 100 times more abundant),
 Iso2HD (the HD$^{16}$O,  HD$^{17}$O and  HD$^{18}$O isotopes are two times more abundant),
 Iso10HD (the HD$^{16}$O,  HD$^{17}$O and  HD$^{18}$O isotopes are ten times more abundant),
 Iso100HD (the HD$^{16}$O,  HD$^{17}$O and  HD$^{18}$O isotopes are 100 times more abundant).

For each case study, we adjusted the abundance of the main isotope to keep a total abundance equal to unity (see \tab{table_isotopes}). Figure~\ref{fig_isotopes} shows that secondary isotopes slightly reduce the height of the overshoot even at a terrestrial abundance (Iso1), but this difference is lower than 2\,W.m$^2$. At higher abundances (Iso2, Iso10, Iso100), the strong absorption of the secondary isotopes reduces the OLR value over the entire studied temperature range. The tendency is the same when only increasing abundances of the HD$^{X}$O isotopes (Iso2HD, Iso10HD, Iso100HD) Nevertheless, \fig{fig_isotopes} shows that increasing the abundance of all the secondary isotopes more efficiently reduces OLR than only increasing HD$^{X}$O abundances (Iso100 and Iso100HD).

\begin{table*}[ht]\footnotesize\centering
\setlength{\doublerulesep}{\arrayrulewidth}
\captionsetup{justification=justified}
\caption{Overview of the isotope abundances per isotope (Isotopes) for the different case studies. The Iso1 case study corresponds to terrestrial abundances (Terr. abun.) while, for the Iso2, Iso10, and Iso100 cases, secondary isotope abundances are multiplied by 2, 10, and 100. For the Iso2HD, Iso10HD, and Iso100HD cases we increase only the abundance of HD$^{X}$O isotopes by multiplying them by a factor of 2, 10, or 100.}
\label{table_isotopes}
\begin{tabular}[c]{cccccccc}
    \hline\hline\hline
    Isotopes  & Terr. abun. & Iso2 & Iso10 & Iso100 & Iso2HD & Iso10HD & Iso100HD  \\
    \hline
    H${_2}^{16}$O & 0.997 & 0.994 & 0.973 & 0.731 & 0.997 & 0.995 & 0.967 \\
    H${_2}^{18}$O & 0.002 & 0.004 & 0.020 & 0.200 & 1.99e-3 & 1.99e-3 & 1.94e-3 \\
    H${_2}^{17}$O & 3.72e-4 & 7.44e-4 & 3.72e-3 & 3.72e-2 & 3.72e-4 & 3.72e-4 & 3.61e-4 \\
    HD$^{16}$O & 3.11e-4 & 6.21e-4 & 3.11e-3 & 3.11e-2 & 6.21e-4 & 3.11e-3 & 3.11e-2 \\
    HD$^{18}$O & 6.23e-7 & 1.25e-6 & 6.23e-6 & 6.23e-5 & 1.25e-6 & 6.23e-6 & 6.23e-5 \\
    HD$^{17}$O & 1.16e-7 & 2.32e-7 & 1.16e-6 & 1.16e-5 & 2.32e-7 & 1.16e-6 & 1.16e-5 \\
    D${_2}^{16}$O & 2.42e-8 & 4.84e-8 & 2.42e-7 & 2.42e-6 & 2.42e-8 & 2.41e-8 & 2.35e-8 \\
    
    \hline\hline\hline
\end{tabular}\normalsize
\end{table*}

\subsubsection{CO$_2$}
\label{co2}

\begin{figure}[h!]
    \centering\includegraphics[width=\linewidth]{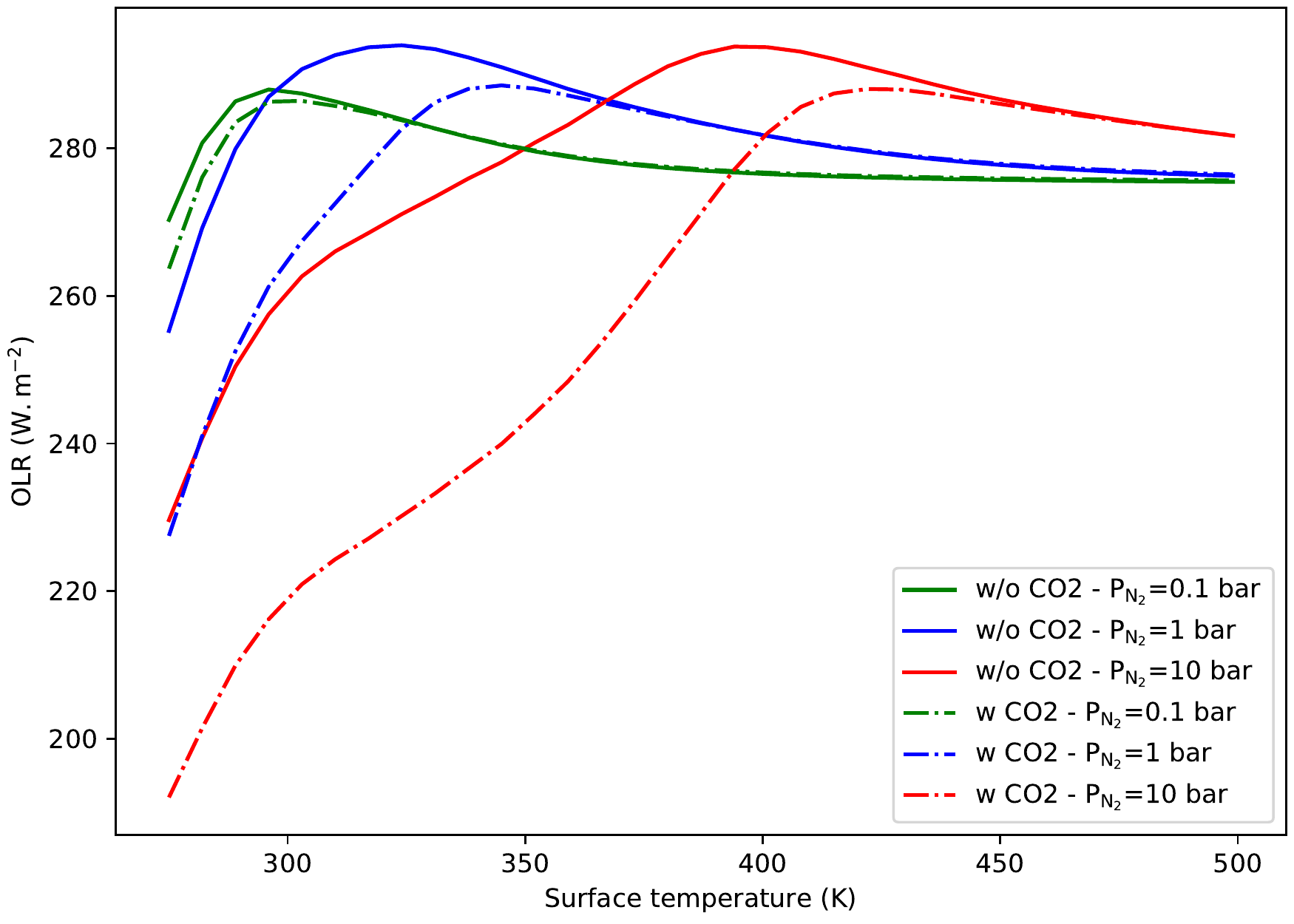}
    \caption{OLR values as a function of the surface temperature with and without CO$_2$ computed with \texttt{Exo\_k}. The full lines are the reference OLR curves proposed in \fig{fig_PyRADS-Conv1D} using \texttt{Exo\_k,} and the dotted lines include 376 ppm of CO$_2$ in the radiative transfer computation.}
    \label{fig_CO2}
\end{figure}

The typical atmospheric compositions of moderately irradiated rocky exoplanets is currently unknown. A common solution to fill this gap is to consider an Earth-like atmosphere. Therefore, several studies added a few parts-per-million of CO$_2$ in the atmospheric composition, such as \cite{kopparapu_habitable_2014}, who added 350~ppm (Ravi K. Kopparapu, personal communication). Such an amount of CO$_2$ can strongly change the OLR values but also the shape of the OLR curve (\fig{fig_CO2}). 
The difference induced by CO$_2$ is higher at low temperatures, where water is not the dominant gas. In other words, for such low temperatures the absorption due to CO$_2$ is non-negligible. 
At high temperatures, the absorption due to water is high enough to overlap the CO$_2$ absorption.
For this reason, if there is CO$_2$ in the simulated atmosphere, it is important to consider it in the analysis to avoid incorrect conclusions. 

In this work, as we explored the effects of nitrogen, we considered an H$_2$O + N$_2$ atmosphere without CO$_2$. 
A few papers \citep[e.g.][]{ramirez_can_2014, popp_transition_2016} discuss the effect of CO$_2$ on the onset of the runaway greenhouse in more detail.
In \fig{fig_CO2}, as the CO$_2$ quantity is a percentage of the N$_2$ pressure (we used 376 ppm of CO$_2,$ as \citealt{leconte_increased_2013}), the amount of CO$_2$ increases when increasing the N$_2$ pressure. This is the reason why the difference between simulations with and without CO$_2$ is stronger at high N$_2$ pressure.
The strong greenhouse power of the CO$_2$ also strongly reduces the height of the overshoot.
Finally, CO$_2$ absorption explains a large part of the gap between the results of \cite{kopparapu_habitable_2014} and those of other.

\subsubsection{Two-stream approximation}
\label{sub_two-stream}

All the radiative transfer models used here employ a two-stream approximation in the infrared where the radiative transport in the atmosphere is described by an upward and a downward flux (\citealt{meador_two-stream_1980}; in fact, as we are focused on the thermal emission in a non-scattering atmosphere, we are only really dealing with one stream).

\begin{figure}[h!]
    \centering\includegraphics[width=\linewidth]{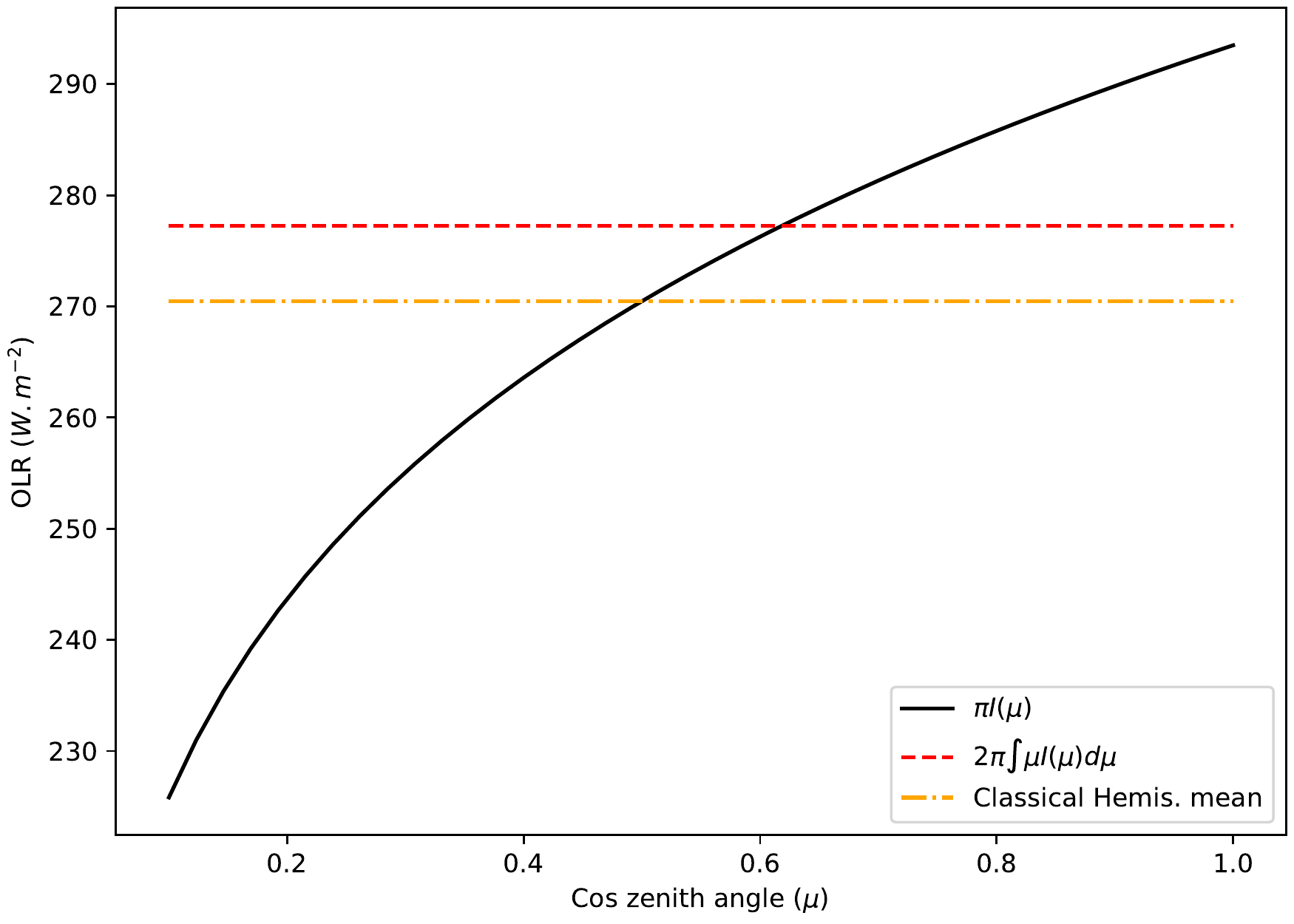}
    \caption{Variation of OLR with respect to the effective zenith angle used to compute the upward intensity (solid curve). The dashed curve shows the OLR computed by properly integrating the intensity over all zenith angles. The horizontal dash-dot curve shows the result of the classical hemispheric mean approximation \citep{toon_rapid_1989}, which systematically underestimates the flux by $\sim$6\,W.m$^{-2}$. The atmosphere model used has T$_\mathrm{surf}$=500\,K and P$_\mathrm{N_2}$=1\,bar.}
    \label{fig_mu_angle}
\end{figure}

Since the seminal work of \citet{toon_rapid_1989}, it is known that, while this approach is numerically efficient, the assumptions made (e.g. the two-stream coefficients used) need to be adapted to the problem at hand to ensure accuracy. For example, as can be seen in \fig{fig_mu_angle}, the classical hemispheric mean approximation (where the intensity is considered constant over each hemisphere) systematically underestimates the outgoing emitted flux compared to a proper integration over the cosine of the zenith angle ($\mu$). This is due to the fact that for a given specific intensity (considered independent of the azimuth angle), $I(\mu)$, the integral for the upward flux,
\begin{align}
F^+ = 2\pi \int_0^1 \mu I(\mu) d \mu,
\end{align}
gives more weight to the rays close to the vertical. Because these rays probe deeper, they have a higher than average intensity (see \fig{fig_mu_angle}, where we show $\pi I(\mu)$ to use the same scale as the flux).
From this figure, we see that the total flux would be better approximated by $\pi I(\mu\sim0.6);$ thus, one might be tempted to use the classical quadrature method and choose $\bar{\mu}\sim0.6$ as the quadrature point. As pointed out by \citet{toon_rapid_1989}, however, this approach leads to a well-known unphysical emissivity that is highly undesirable if one wants to conserve energy. 

To avoid having to use an expensive angular integration while preserving an accurate and physical solution, the \texttt{Exo\_k}-RT method (\tab{table_models}) uses the following approximation in the infrared, which is a variation on the hemispheric approximation. Following notations from \citet{toon_rapid_1989}, the equation for the upward flux (and similarly for the downward flux, $F^-$) writes
\begin{align}
\label{2stream1}
\factor\frac{\partial F^+}{\partial \tau}&= \gamma_1 F^+ - \gamma_2 F^- - 2\pi (1-\omega_0) B,
\end{align}
where $\gamma_1=2-\omega_0(1+g)$ and $\gamma_2=\omega_0(1-g)$, $\omega_0$ and $g$ are the single scattering albedo and asymmetry factor of the medium, and $B$ is the Planck function. The difference with the classical hemispheric approximation is the factor $\factor$ that accounts for the fact that, when assuming the hemi-isotropy of the intensity, choosing an \textit{\emph{effective}} optical depth ($d \tau /\factor$) that is different from the vertical one does not introduce a greater level of approximation.

Interestingly, choosing an effective zenith angle whose cosine is $\bar{\mu}$ and defining $\factor\equiv 2\bar{\mu}$ \footnote{The factor of 2 comes from the fact that the hemispheric-mean approximation can be seen as looking at a ray with $\bar{\mu}=1/2$, so that $\factor$ can be seen as a correction factor accounting for the departure from the hemispheric-mean effective zenith angle.}, \eq{2stream1} rewrites
\begin{align}
\label{2stream2}
\frac{\partial F^+}{\partial \tau}&= \left(\frac{2-\omega_0(1+g)}{2\bar{\mu}}\right) F^+ - \left(\frac{\omega_0(1-g)}{2\bar{\mu}}\right) F^- - \frac{\pi}{\bar{\mu}} (1-\omega_0) B.
\end{align}
By comparison with Table 1 of \citet{toon_rapid_1989}, one can directly see that this assumption is completely equivalent to the quadrature approximation where an arbitrary quadrature point $\bar{\mu}$ is chosen instead of the Gauss-Legendre choice (1/$\sqrt{3}$), except for the change to the source term. This change is what makes our approximation retain the physical consistency of the hemispheric mean approach as $F^+ \rightarrow \pi B$ in an opaque, non-scattering medium, whatever $\bar{\mu}$ we choose. 

Our modified hemispheric-mean approximation, which is the default method in \texttt{Exo\_k}, thus combines the energy-conserving feature of the classical hemispheric-mean with the flexibility of choosing the best effective zenith angle for a given problem. In practice it is easy to adapt any numerical implementation of the classical two-stream approach to solve this system: i) use the $\gamma$ coefficients from \eq{2stream2} and ii) change the $2\pi\mu_1$ coefficient in Eq.\,(27) of \citet{toon_rapid_1989} into $\pi$ to enforce energy conservation.

In the context of a non-scattering atmosphere, it is easy to show that our modified hemispheric-mean approximation is equivalent to computing the intensity for a given $\bar{\mu}$ and assuming $F^+=\pi I(\bar{\mu})$. As can be seen in \fig{fig_mu_angle} and is discussed above, $\bar{\mu}=0.6$ yields a less than 1\,W.m$^{-2}$  difference over the whole range of atmospheres modelled here. However, we highlight that care should be given to the radiative transfer method used when comparing model results as different methods can lead to differences of $\sim 6$\,W/m$^2$. These differences are visible between kcm1d (which uses the classical hemispheric mean approximation) and PyRADS (which assumes $F^+=\pi I(\bar{\mu})$ with $\bar{\mu}=0.6$) in \fig{model_comparison}.

\subsubsection{Convection scheme}

\begin{figure}[h!]
    \centering\includegraphics[width=\linewidth]{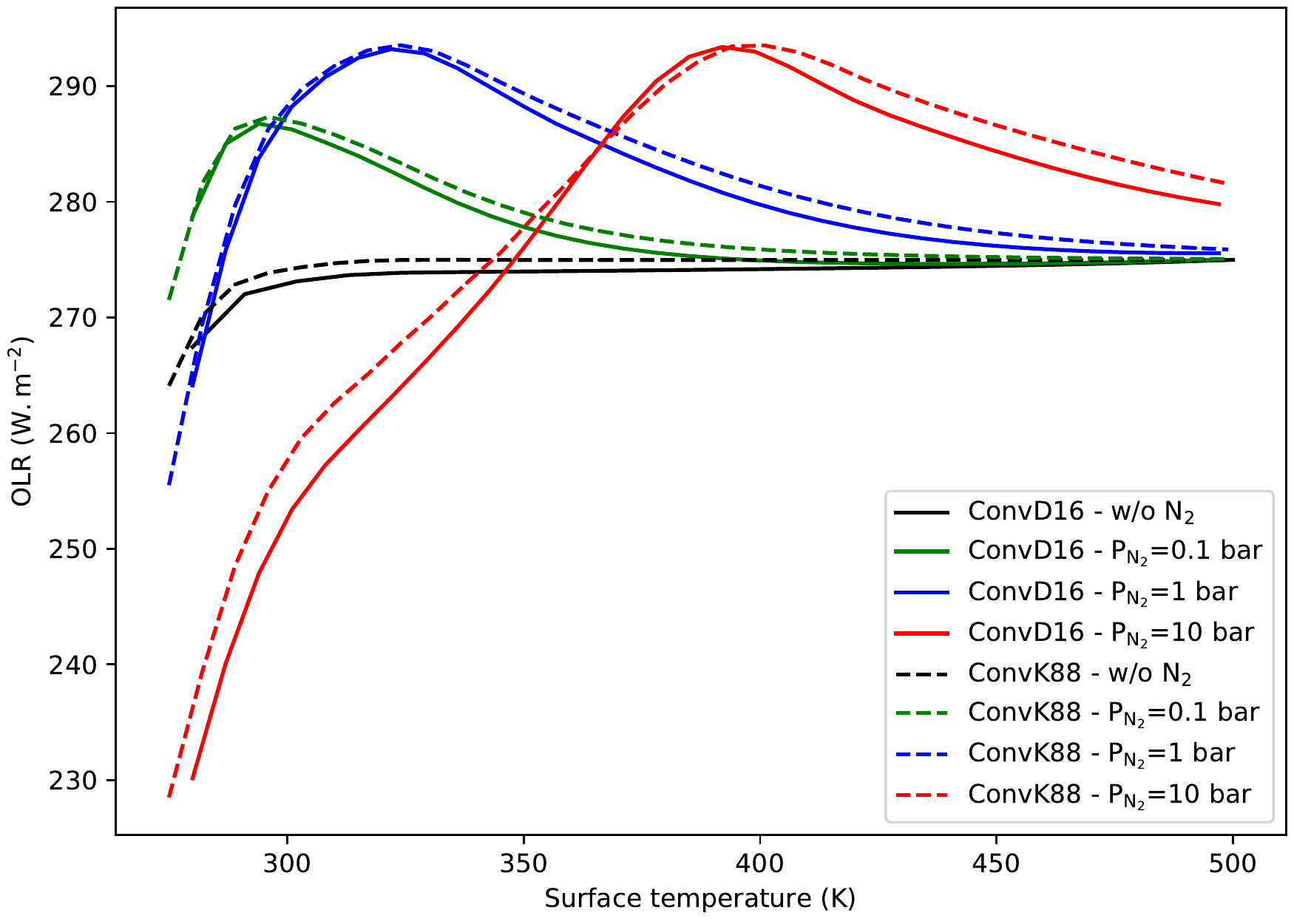}
    \caption{OLR values as a function of the surface temperature for different nitrogen pressures assuming the Conv\_K88 or the Conv\_D16 scheme using PyRADS-Conv1D.}
    \label{fig_conv}
\end{figure}

The two convection schemes used in this work and presented in \sect{sub_conv_schemes} are based on the two main adiabatic lapse rates available in the literature. In \app{sub_adiabat}, we show that the difference between them comes from the definition of the entropy of the condensable gas. 
As explained in \sect{sub_conv_schemes}, we used the Conv\_K88 scheme as reference because it is based on look-up tables, which is more accurate than the perfect gas laws used in the Conv\_D16 scheme. It is interesting to quantify the uncertainty bewteen OLR curves computed using one scheme or the other. 
We modified PyRADS-Conv1D to use the Conv\_K88 of the Conv\_D16 scheme and we obtained the curves presented in \fig{fig_conv}. We show that the difference is smaller than 5\,W.m$^2$. We also notice that the Conv\_D16 scheme systematically underestimates the OLR, probably mainly by overestimating the water content.

\subsection{Sensitivity studies on numerical setup}
\label{sub_num_study}
The second set of sensitivity tests is focused on numerical parametrisations.

\subsubsection{Vertical grid resolution}
\label{sub_vertical_grid}
\begin{figure}[h!]
    \centering\includegraphics[width=\linewidth]{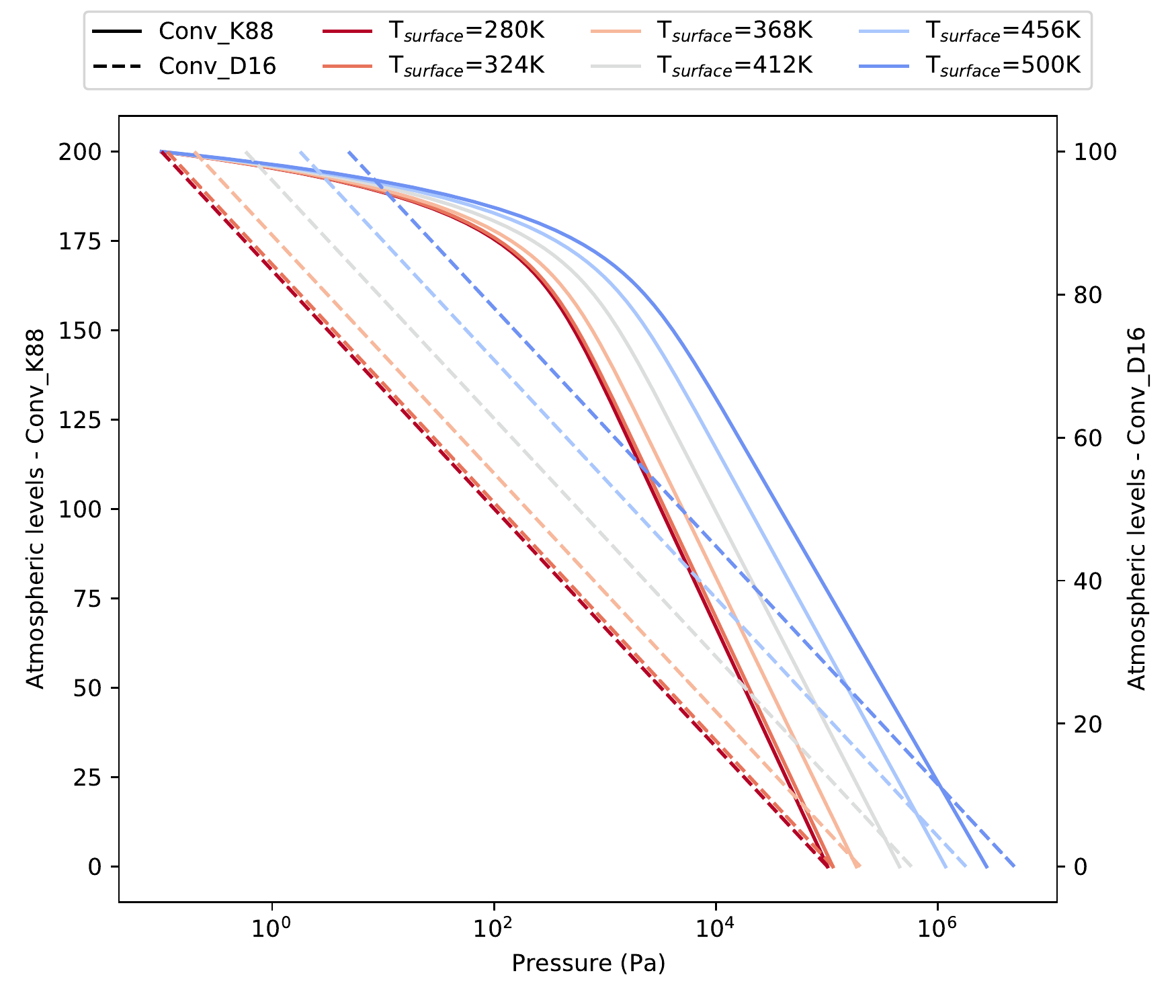}
    \caption{Comparison of pressure distribution over the vertical grid of the model for the Conv\_K88 and the Conv\_D16 methods. Here, P$_{\textrm{N}_2}$=1~bar.}
    \label{fig_atm_levels}
\end{figure}

Aside from using two different adiabats, the convective schemes used in this work assume two different definitions of the vertical grid.
The Conv\_D16 scheme provides a logarithmic distribution of the levels with a fixed difference between the surface pressure and pressure at the top of the atmosphere, while the Conv\_K88 scheme adapts the height of the levels relatively to the amount of water from the surface up to 0.1~Pa (\fig{fig_atm_levels}).
In other words, in the Conv\_K88 scheme the wet atmospheric levels near the surface are thinner than the drier ones at the top of the atmosphere. The aim is to increase the vertical resolution of the 'radiatively active' part of the atmosphere. 
The number of vertical levels required to compute an accurate value of the OLR does not depend on the nitrogen pressure or on the surface pressure, but this number is higher than for the Conv\_D16 scheme. Consequently, using the Conv\_K88 scheme is more accurate even if the computation time is more expensive. For this reason, we used it in PyRADS-Conv1D.
A convergence curve of the OLR relative to the number of vertical levels highlights that 200 levels are needed for the Conv\_K88 scheme to keep an error smaller than 1W/m$^2$ compared to the converged OLR value.
We assumed a stratospheric temperature equal to 200~K according to \cite{kasting_runaway_1988}. That work showed that a value lower than 250~K does not influence the OLR because for such low stratospheric temperatures the vapour pressure is radiatively negligible in the upper layers of the atmosphere (by assuming a pseudo-adiabatic approximation).

In the Conv\_D16 scheme, the vertical grid does not depend only on the number of levels but also on the difference in pressure between the top and the bottom of the atmosphere. This induces a dependency on the number of levels on the surface temperature as shown in \tab{table_vertical_levels}.
Indeed, because of the pseudo-adiabatic approximation the total surface pressure increases strongly between 280~K and 500~K ($\approx$1~bar to $\approx$26~bar if P$_{\textrm{N}_2}$=1~bar), and the difference between pressures at the top and at the bottom of the atmosphere may be high to accurately represent the whole atmosphere. 
Moreover, increasing the nitrogen pressure reduces the height of the radiatively active part of the atmosphere by drying it up; thus, the required number of levels increases. 
We also observed that using a stratospheric temperature equal to 150~K (and not 200~K as in the Conv\_K88 scheme) allows us to slightly reduce the required number of levels.
Finally, according to Table~\ref{table_vertical_levels}, in the Conv\_D16 scheme we assume 100 vertical levels and a difference between the pressure at the bottom and at the top of the atmosphere equal to 10$^6$~Pa. This setup provides a good compromise between accuracy and computation time with an error at most equal to 1~W/m$^2$ compared to the converged value.

\begin{table}[ht]\footnotesize\centering
\setlength{\doublerulesep}{\arrayrulewidth}
\captionsetup{justification=justified}
\caption{Overview of the number of atmospheric levels needed in the Conv\_D16 scheme relatively to the nitrogen pressure, the surface temperature (Tsurf), the stratospheric temperature (Tstrat), and the difference between the pressure at the bottom and at the top of the atmosphere (Pres. diff.).}
\label{table_vertical_levels}
\begin{tabular}[c]{ccccc}
    \hline\hline\hline
    Nitrogen pressure  & Tsurf & Tstrat & Pres. diff. & Nb. of levels  \\
    \hline
    0.1~bar & 300~K & 150~K & 10$^4$~Pa & 30 \\
    1~bar & 300~K & 150~K & 10$^5$~Pa & 100 \\
    10~bar & 300~K & 150~K & 10$^6$~Pa & 200 \\
    \hline
    0.1~bar & 500~K & 150~K & 10$^7$~Pa & 60 \\
    1~bar & 500~K & 150~K & 10$^7$~Pa & 60 \\
    10~bar & 500~K & 150~K & 10$^7$~Pa & 60 \\
    \hline\hline\hline
\end{tabular}\normalsize
\end{table}

\subsubsection{Spectral resolution}

\begin{figure}[h!]
    \centering\includegraphics[width=\linewidth]{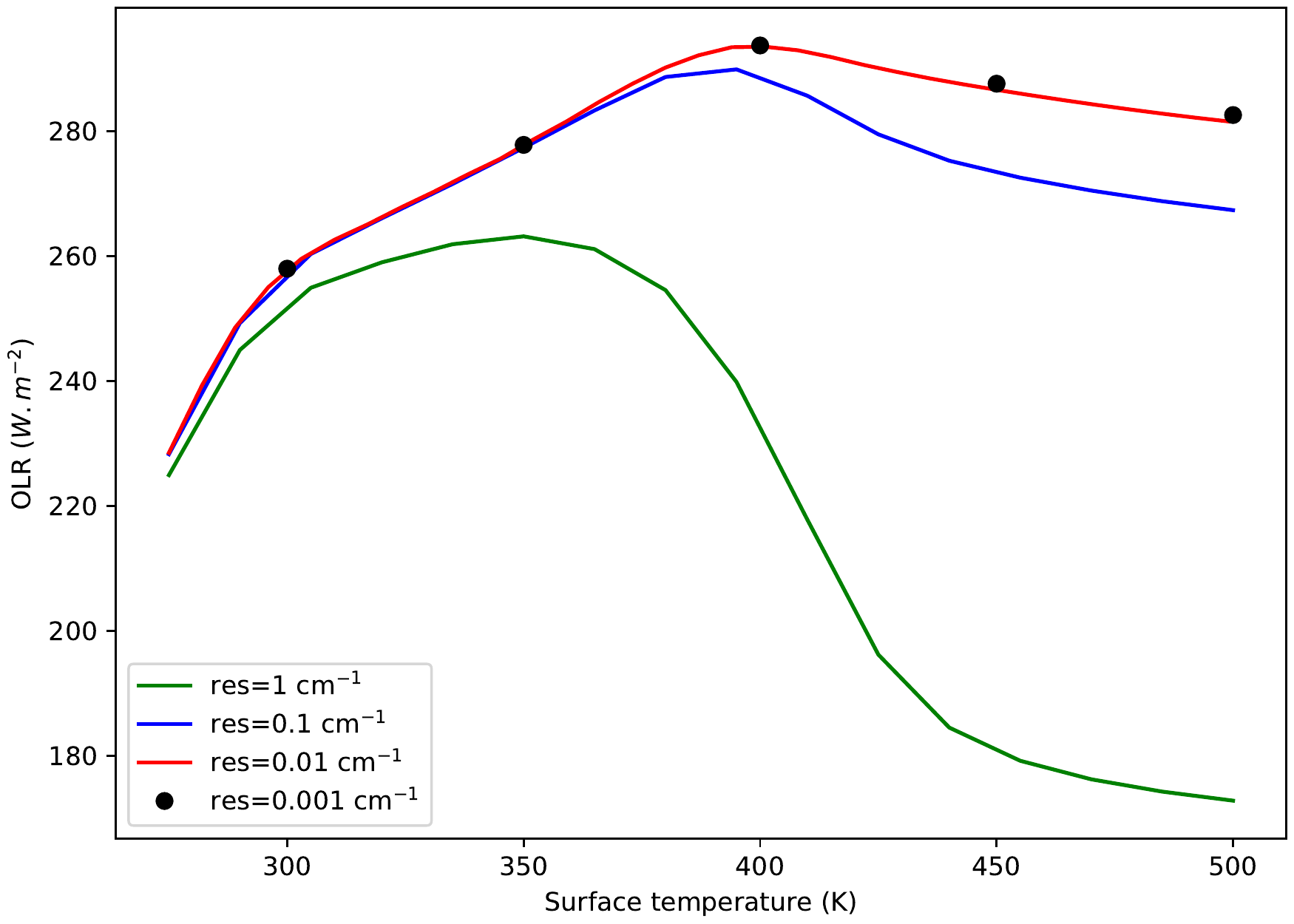}
    \caption{OLR values as a function of the surface temperature for different spectral resolution with P$_{\textrm{N}_2}$=10~bar and using PyRADS-Conv1D.}
    \label{fig_spectral_reso_olr}
\end{figure}

\begin{figure}[h!]
    \centering\includegraphics[width=\linewidth]{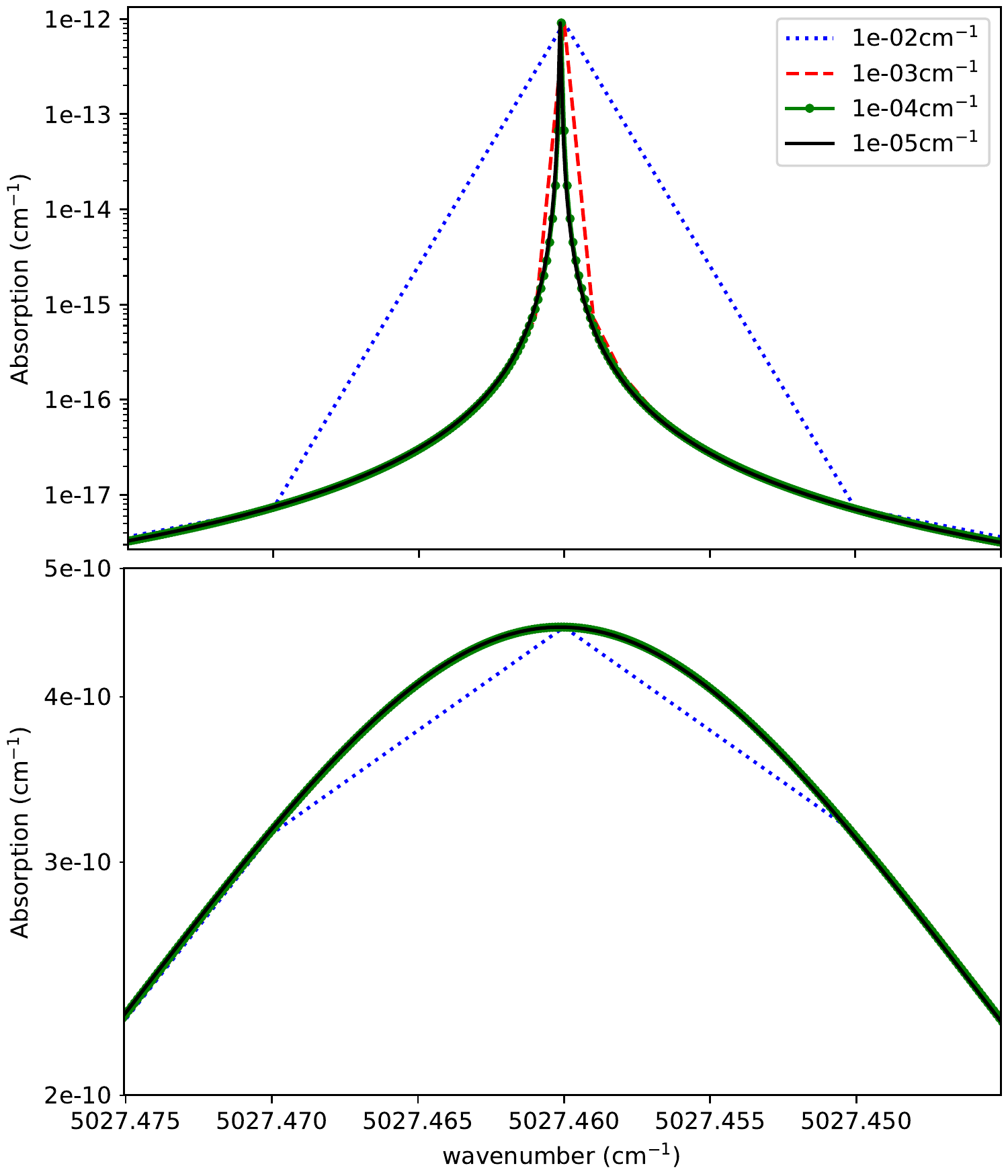}
    \caption{Absorption line at 5027.460050~cm$^{-1}$ for different spectral resolutions and using a Lorentz profile. The top panel is the absorption for P=6~Pa and T=220~K, and the bottom panel is the absorption
for P=3500~Pa and T=300~K.}
    \label{fig_spectral_reso}
\end{figure}

For the line-by-line method, as we assume a Lorentz profile for the absorption lines, the spectral resolution needs to be high enough to accurately reproduce the real spectrum.
The line-by-line method needs a minimal spectral resolution of 0.01 cm$^{-1}$ to give an accurate value of the OLR (\fig{fig_spectral_reso_olr}) as shown by \cite{koll_earths_2018}. The absorption lines are not exactly resolved, as is visible in \fig{fig_spectral_reso}, but the cumulative error on the entire spectrum is negligible. 
Producing a correlated-k table requires high-resolution spectra; therefore, we used a spectral resolution of 0.001 cm$^{-1}$ , which is able to resolve each absorption line.\\
\fig{fig_spectral_reso} shows that by increasing the pressure and the temperature, the width of the line becomes larger; thus, a lower resolution is sufficient. Nevertheless, in the highest atmospheric levels the pressure is low, and a high spectral resolution is always required.

\subsubsection{Correlated-k-mixing ratio grid}
Because of the non-usual transition between an N$_2$ -dominated and an H$_2$O atmosphere, we tested the resolution of the volume-mixing ratio (vmr) grid of the correlated-k table. The original resolution of the vmr grid ($Q=\{1\textrm{e}^{-6}, 1\textrm{e}^{-5}, 1\textrm{e}^{-4}, 1\textrm{e}^{-3}, 1\textrm{e}^{-2}, 1\textrm{e}^{-1}, 1\}$) was increased to $Q=\{1\textrm{e}^{-6}, 1\textrm{e}^{-5}, 1\textrm{e}^{-4}, 1\textrm{e}^{-3}, 1\textrm{e}^{-2}, 5\textrm{e}^{-2}, 1\textrm{e}^{-1}, 5\textrm{e}^{-1}, 1\},$ but it does not significantly change the OLR values.

\subsubsection{Correlated-k-interpolation scheme}
\label{sub_interpolation}

Correlated-k tables contain absorption spectra for given pressure, temperature, and vmr grids. To be able to compute the absorption of each atmospheric level, the models need to interpolate these tables on different pressure, temperature, and vmr values.
Because of the wide diversity of interpolation schemes, we do not describe them in detail. However, by using two different schemes (from two different models), we caveat that a part of the differences highlighted in \fig{model_comparison} can come from the choice of the correlated-k-interpolation scheme.

\begin{figure}[h!]
    \centering\includegraphics[width=\linewidth]{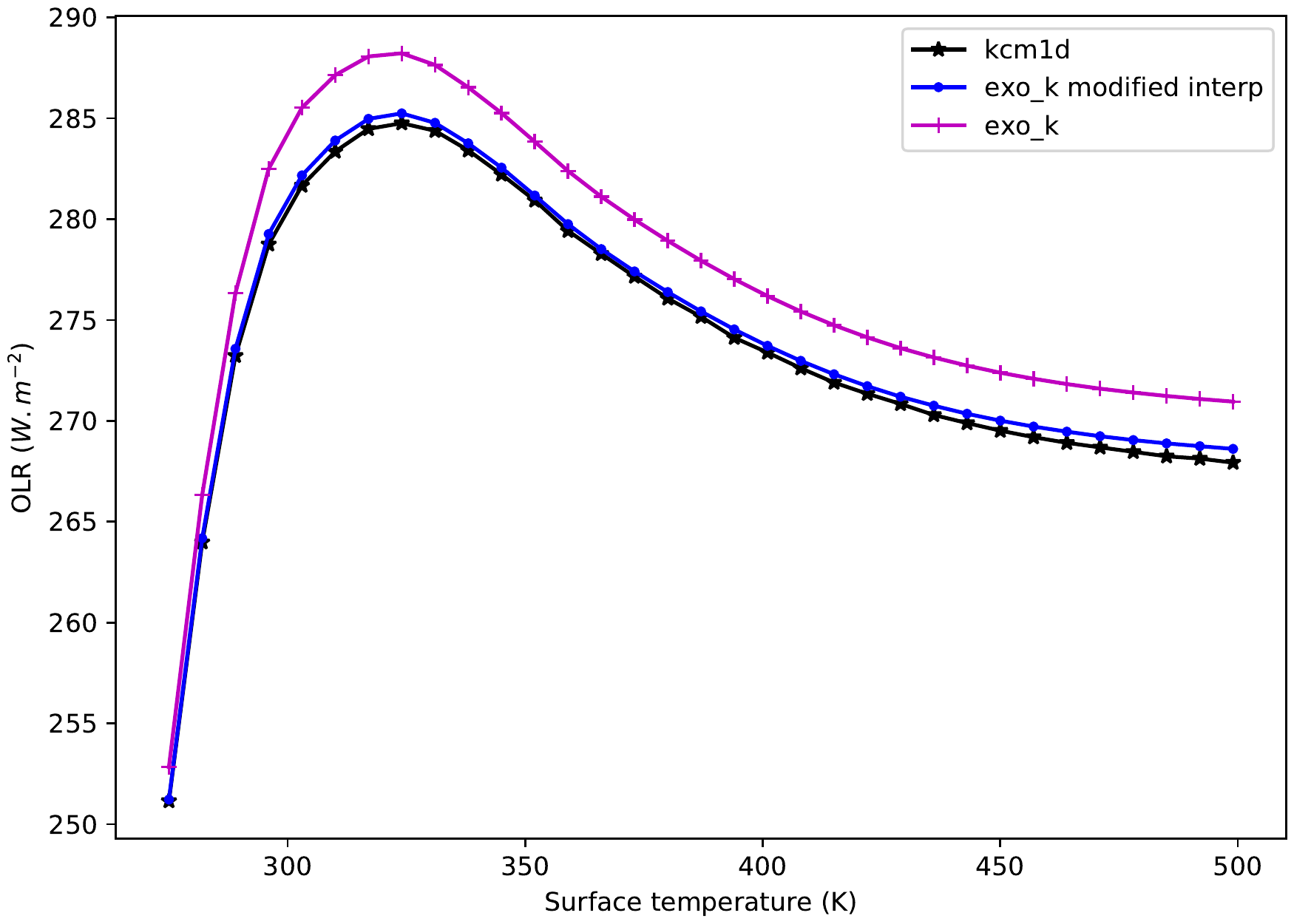}
    \caption{OLR as a function of the surface temperature for different interpolation schemes: kcm1d (black line) and \texttt{Exo\_k} (magenta line). The blue line corresponds to \texttt{Exo\_k} using the kcm1d interpolation scheme. Here, P$_{\textrm{N}_2}$=1~bar.}
    \label{fig_interp}
\end{figure}

The two correlated-k radiative transfer models we used in this work (\texttt{Exo\_k} and kcm1d in \tab{table_models}) have different interpolation schemes and produce different OLR curves for the same correlated-k table. Nevertheless, this difference is smaller than 3~W.m$^2$. 
We were able to reproduce OLR curves from kcm1d with a modified version of \texttt{Exo\_k} that uses the same interpolation scheme as kcm1d, with an error lower than 1~W.m$^2$. This proves that the only difference in \fig{fig_interp} comes from the interpolation scheme.
As kcm1d is derived from the LMD-Generic model, it inherits an interpolation scheme optimised for such 3D models; thus, this difference of 3~W.m$^2$ is a second-order error regarding other uncertainties of GCMs.

Although the detail of the interpolation of \texttt{Exo\_k} and kcm1d are different, the global scheme is the same.
The absorption spectrum (K) is interpolated on the pressure (P), temperature (T), and vmr (x) as follows:
$$\mathrm{\ln K=fct(\ln P, T, \ln x).}$$
This global scheme provides the most accurate interpolated spectra, which makes sense regarding the logarithmic distribution of the pressure and vmr grids of the correlated-k tables.

\section{Adiabatic lapse rate formulations}
\label{sub_adiabat}
One-dimensional radiative-convective climate models usually use a pseudo-adiabatic lapse rate (also referred to as adiabat) to compute the temperature profile of the atmosphere. There are two widely used formulations of this adiabat in the literature for a moist atmosphere. The first one is proposed by \cite{kasting_runaway_1988} based on \cite{ingersoll_runaway_1969}:
\begin{equation} \label{eq_kasting88}
    \frac{d\ln P}{d\ln T}=\frac{P_c}{P}\frac{d\ln P_c}{d\ln T}+\frac{P_n}{P}\left(1+\frac{d\ln \rho_c}{d\ln T}-\frac{d\ln\alpha_c}{d\ln T}\right)
,\end{equation}
with
\begin{equation} \label{eq_kasting88_2}
    \frac{d\ln\alpha_c}{d\ln T}=\frac{\frac{R^*}{M_n}\left(\frac{d(ln \rho_c)}{d\ln T}\right)-c_{vn}-\alpha_c\frac{ds_c}{d\ln T}-\alpha_l\frac{ds_l}{d\ln T}}{\alpha_c(s_c-s_l)+\frac{R^*}{M_n}}
.\end{equation}
The second one is proposed by \cite{ding_convection_2016} based on the dry adiabat of \cite{pierrehumbert_principles_2010}:
\begin{equation}\label{eq_pierrehumbert16}
    \frac{d\ln P}{d\ln T}=\frac{P_c}{P}\frac{L}{R_cT}+\frac{P_n}{P}\frac{c_{pn}}{R_n}\times\frac{1+\left(\frac{c_{pc}}{c_{pn}}+\left(\frac{L}{R_cT}-1\right)\frac{L}{c_{pn}T}\right)\alpha_c}{1+\frac{L}{R_nT}\alpha_c}
.\end{equation}

Both moist lapse rates presented above are valid for a non-dilute atmosphere (i.e. for an atmosphere where water is a trace gas). \cite{leconte_increased_2013, leconte_condensation-inhibited_2017} proposed a third formulation equivalent - but not strictly equal - to \cite{ding_convection_2016} (see \eq{eq_leconte13}). Nevertheless, \cite{leconte_condensation-inhibited_2017} highlighted an  interesting molecular mass gradient effect.

Through a few mathematical reformulations of the adiabat proposed by \cite{leconte_increased_2013}, we obtain the following equation, which is strictly equal to \eq{eq_pierrehumbert16} under the pseudo-adiabatic approximation (i.e. $\alpha_l=0$):
\begin{equation} 
\label{eq_leconte13}
    \frac{d\ln P}{d\ln T}=\frac{P_c}{P}\frac{L(T)}{R_cT}+\frac{P_n}{P}\frac{c_{pn}}{R_n}\times\frac{1+\left(\frac{c_{pc}}{c_{pn}}+\left(\frac{L}{R_cT}-1\right)\frac{L}{c_{pn}T}\right)\alpha_c+\frac{c_{pl}}{c_{pn}}\alpha_l} {1+\frac{L}{R_nT}\alpha_c}
.\end{equation}

Indices of variables in the \eqsss{eq_pierrehumbert16}{eq_kasting88}{eq_kasting88_2}{eq_leconte13} indicate what gas is considered: $c$ for the condensable gas, $n$ for the non-condensable gas, and $l$ for the condensed gas. Here, $P$ is the pressure, $T$ the temperature, $M$ the molecular weight, $\rho$ the density, $R^*$ the perfect gas constant, $R$ the specific gas constant, $\alpha_i=m_i/m_n$ the mass-mixing ratio per unit of non-condensable gas ($m_n$) for the species $i$, $L$ the latent heat, and $c_p$ the heat capacity.\\

These equations are different but can be equalised through three simple equations. The main difference between \eqs{eq_pierrehumbert16}{eq_kasting88} comes from the initial definition of the entropy. Indeed, \eq{eq_pierrehumbert16} defines the entropy of the condensable gas as a perfect gas entropy. 
Equation\,(\ref{eq_kasting88}) does not state the entropy of the condensable gas. Therefore, it is possible to use values from experiments for the condensable entropy to reduce inaccurate results at high temperatures (i.e. near the critical point).
In both cases, the total gas pressure is defined using the perfect gas law.

\subsection{Development of the adiabatic lapse rate of \cite{ding_convection_2016}}
We start with the definition of the total heat budget per unit of mass $\delta Q_m$. Here, we consider a volume per unit of mass ($dV/m$):
\begin{equation}  \label{eq_deltaQ}
\delta Q_m=c_vdT+Pd\rho^{-1}=c_pdT-\rho^{-1}dP
,\end{equation} 
where $Pd\rho^{-1}\leftrightarrow P\frac{dV}{m}$ and $c_p=c_v+\frac{R^*}{M}$. \\
Using \eq{eq_deltaQ}, we define the total heat budget $\delta Q$ of a gas mixture \citep{pierrehumbert_principles_2010}:
$$\delta Q=(m_n+m_c)\delta Q_m=m_nc_{pn}dT-\frac{m_n}{\rho_n}dP_n+m_cc_{pc}dT-\frac{m_c}{\rho_c}dP_c+Ldm_c,$$
where $\delta Q$ is the total heat budget and $\delta Q_m$ is the total heat budget per unit of mass.
We consider a pseudo-adiabatic system, and thus all the condensation is immediately removed by precipitation. In other words, the pressure of the condensable gas is equal to the saturation pressure.

By dividing the previous equation by the mass of the non-condensable gas ($m_n$) we obtain the total heat budget per unit of non-condensable gas:
\begin{multline}
    \delta Q_{m_n}=(1+\alpha_c)\delta Q_m=\left(c_{pn}dT+R_nT\frac{dP_n}{P_n}\right)\\
    +\left(c_{pc}\alpha_c dT -\alpha_c R_c T\frac{dP_c}{P_c}\right)+Ld\alpha_c,
\end{multline}
where $\alpha_c=m_c/m_n$.
By dividing the previous equation by the temperature then by rewriting it, we obtain
\begin{multline}
    \frac{\delta Q_{m_n}}{T}=\left(c_{pn}\left(c_{pc}+\left(\frac{L}{R_cT}-1\right)\frac{L}{T}\right)\alpha_c\right)d\ln T- \\
    \left(1+\frac{L}{R_nT}\alpha_c\right)R_n~d\ln P_n.
\end{multline}
By definition of an adiabatic (or pseudo-adiabatic) transformation, $\delta Q=0$. Therefore, it is possible to define the dry adiabatic lapse rate \citep{pierrehumbert_principles_2010}:
\begin{equation} \label{eq_dlnPn}
    \frac{d\ln P_n}{d\ln T}=\frac{c_{pn}}{R_n}\times\frac{1+\left(\frac{c_{pc}}{c_{pn}}+\left(\frac{L}{R_cT}-1\right)\frac{L}{c_{pn}}T\right)\alpha_c}{1+\frac{L}{R_nT}\alpha_c}
.\end{equation}
By assuming that the total pressure is the sum of both condensable and non-condensable pressures, we obtain
\begin{equation} \label{eq_dlnP}
    \frac{d\ln P}{d\ln T}=\frac{P_c}{P}\frac{d\ln P_c}{d\ln T}+\frac{P_n}{P}\frac{d\ln P_n}{d\ln T}
.\end{equation}
Then, by using \eqs{eq_dlnPn}{eq_dlnP} and the equation of Clausius-Clapeyron (\eq{eq_clausius}), 
\begin{equation} \label{eq_clausius}
    \frac{d\ln P_c}{d\ln T}=\frac{L}{R_cT}
.\end{equation}
We obtain the moist adiabatic lapse rate proposed by  \cite{ding_convection_2016} (\eq{eq_pierrehumbert16}). We assume that $c_p$ does not depend on temperature, but this assumption becomes false near the critical point.

\subsection{Development of the adiabatic lapse rate of \cite{kasting_runaway_1988}}
The moist adiabatic lapse rate proposed by \cite{kasting_runaway_1988} uses equations from \cite{ingersoll_runaway_1969}.
First, \cite{ingersoll_runaway_1969} assumed the entropy of an adiabatic system $S$, per unit of mass of non-condensable gas:
\begin{equation}\label{eq_S}
S=s_n+\alpha_cs_c+\alpha_ls_l
,\end{equation}
where $s_n$, $s_c,$ and $s_l$ are the entropies of the non-condensable, the condensable, and the liquid part, and $\alpha_X=\frac{\rho_X}{\rho_n}$. Near the critical point, it may not be valid to treat each constituent separately.
By rewriting \eq{eq_S}, we obtain
\begin{equation}\label{eq_dS}
\frac{dS}{dT}=\frac{\partial s_n}{\partial T}+\frac{\partial s_n}{\partial \rho_n}\frac{d\rho_n}{dT}+\alpha_c\frac{ds_c}{dT}+s_c\frac{d\alpha_c}{dT}+\alpha_l\frac{ds_l}{dT}+s_l\frac{d\alpha_l}{dT}
.\end{equation}
The law of mass conservation during an adiabatic expansion provides the following equality: $$\frac{d\alpha_l}{dT}=-\frac{d\alpha_c}{dT,}$$
and, using $\rho_n=\frac{\rho_c}{\alpha_c}$,
$$\frac{d\rho_n}{dT} = \frac{1}{\alpha_c}\frac{d\rho_c}{dT}-\frac{\rho_c}{\alpha_c^2}\frac{d\alpha_c}{dT.}$$
Thanks to these equalities, we can rewrite \eq{eq_dS}:
$$\frac{dS}{dT}=\frac{\partial s_n}{\partial T}+\frac{1}{\alpha_c}\frac{\partial s_n}{\partial \rho_n}\left(\frac{d\rho_c}{dT}-\frac{\rho_c}{\alpha_c}\frac{d\alpha_c}{dT}\right)+\alpha_c\frac{ds_c}{dT}+\alpha_l\frac{ds_l}{dT}+(s_c-s_l)\frac{d\alpha_c}{dT.}$$
If the non condensable gas behaves as an ideal gas, we can assume $T\frac{\partial s_n}{\partial T}=\frac{c_{vn}}{M_n}$ and $\rho_n\frac{\partial s_n}{\partial \rho_n}=-\frac{R^*}{M_n}$; thus,
\begin{equation}\label{eq_dS_final}
\frac{dS}{dT}=\frac{c_{vn}}{TM_n}-\frac{R^*d\rho_c}{M_n\rho_cdT}+\frac{R^*}{\alpha_cM_n}\frac{d\alpha_c}{dT}+\alpha_c\frac{ds_c}{dT}+\alpha_l\frac{ds_l}{dT}+(s_c-s_l)\frac{d\alpha_c}{dT}
.\end{equation}
From \eq{eq_dS_final} by assuming the adiabatic hypothesis ($dS=0$), we obtain \eq{eq_dav} \citep{ingersoll_runaway_1969} and \eq{eq_dlnav} \citep{kasting_runaway_1988}:
\begin{equation} \label{eq_dav}
    \frac{d\alpha_c}{dT}=\frac{\frac{R^*}{M_n}\frac{d\ln \rho_c}{dT}-\frac{c_{vn}}{TM_n}-\alpha_c\frac{ds_c}{dT}-\alpha_l\frac{ds_l}{dT}}{s_c-s_l+\frac{R^*}{\alpha_cM_n}}
,\end{equation}

\begin{equation} \label{eq_dlnav}
    \frac{d\ln\alpha_c}{d\ln T}=\frac{\frac{R^*}{M_n}\left(\frac{d(\ln \rho_c)}{d\ln T}\right)-c_{vn}-\alpha_c\frac{ds_c}{d\ln T}-\alpha_l\frac{ds_l}{d\ln T}}{\alpha_c(s_c-s_l)+\frac{R^*}{M_n}}
.\end{equation}
As in \cite{kasting_runaway_1988}, from now we assume a pseudo-adiabatic expansion; hence, $\alpha_l=0$.
By assuming that the total pressure is the sum of both condensable and non-condensable pressures, and using the perfect gas law, we obtain
\begin{equation} \label{eq_dlnP_K88}
P = \frac{R^*}{M_n}\rho_nT + \frac{R^*}{\beta M_c}\rho_cT
,\end{equation}
where $\beta=\beta(\rho_c,T)$ is a parameter that expresses the degree to which the condensable gas departs from ideality. 
By rewriting \eq{eq_dlnP_K88} and using \eq{eq_dlnav}, we obtain the adiabatic lapse rate proposed by \cite{kasting_runaway_1988}:
\begin{equation} \label{eq_Kasting}
    \frac{d\ln P}{d\ln T}=\frac{P_c}{P}\frac{d\ln P_c}{d\ln T}+\frac{P_n}{P}\left(1+\frac{d\ln \rho_c}{d\ln T}-\frac{d\ln\alpha_c}{d\ln T}\right)
.\end{equation}
Here, the entropy of the condensable part ($s_c$) is explicitly defined in the equation. In the Conv\_K88 scheme, which uses \cite{kasting_runaway_1988} equations, we compute the water entropy thanks to experimental look-up tables \citep{haar_nbsnrc_1984}.
\subsection{Equalisation of the adiabatic lapse rates}
The two formulations of the adiabatic lapse rate (\eqs{eq_pierrehumbert16}{eq_kasting88}) can be equalised assuming the following assumptions:
\begin{enumerate}[label=\Alph*.]
    \item $(s_c-s_l) = \frac{L}{T}$
    \item $\frac{ds_c}{d\ln T} = c_{pc}-\frac{L}{T}$
    \item $\frac{d\ln \rho_c}{d\ln T} = \frac{L}{R_cT}-1$
\end{enumerate}
By definition, equality A is always verified and corresponds to the entropy of phase change.
Equalities B and C are only verified if we use the equation of Clausius-Clapeyron (Eq.~\ref{eq_clausius}) to link vapour pressure, temperature, and latent heat.

We show here that the main difference between adiabatic lapse rates from \cite{kasting_runaway_1988} and \cite{ding_convection_2016} comes from the definition of the entropy of the condensable gas. Indeed, \eq{eq_pierrehumbert16} \citep{ding_convection_2016} implicitly uses the definition of the entropy of a perfect gas \citep{pierrehumbert_principles_2010}, which becomes incorrect at high temperatures (i.e. near the critical point). The second difference is the use of the Clausius-Clapeyron equation (\eq{eq_clausius}), which is an approximation of the more general Clausius equation. This Clausius-Clapeyron equation is only valid for a perfect gas far from the critical point.
For these reasons, we choose to use the Conv\_K88 method based on the moist adiabatic lapse rate from \cite{kasting_runaway_1988} for PyRADS-Conv1D, used as a reference in this paper.

\end{document}